\documentclass[traditabstract]{aa}
\usepackage{txfonts}
\usepackage{graphicx}
\usepackage{lscape}
\def\ltsima{$\; \buildrel < \over \sim \;$}
\def\simlt{\lower.5ex\hbox{\ltsima}}
\def\gtsima{$\; \buildrel > \over \sim \;$}
\def\simgt{\lower.5ex\hbox{\gtsima}}
\def\cgs{{erg cm$^{-2}$ s$^{-1}$}}
\def\ergs{{erg s$^{-1}$}}
\def\cm2{{cm$^{-2}$}}

\def\fhx{{$F_{\rm 2-10}$}}

\def\lum{{$L_{2-10}$}}
\def\lumo{{$L_{2-10}^{Obs}$}}
\def\p1{{Paper I}}

\def\xmm{{\em XMM--Newton}}
\def\chandra~{{\em Chandra}}

\def\nhz{{N$_{\rm H}^{\rm z}$}}
\def\chandra{{\em Chandra}}

\def\xmm{{\em XMM--Newton}}

\def\nh{{N$_{\rm H}$}}
\def\epic{{\em EPIC}}
\def\mosuno{{\em MOS1}}
\def\mosdue{{\em MOS2}}
\def\mos{{\em MOS}}
\def\pn{{\em PN}}
\def\xr{{X--ray}}
\def\f14{{10$^{-14}$}}
\def\f13{{10$^{-13}$}}
\def\f12{{10$^{-12}$}}
\def\f11{{10$^{-11}$}}
\def\e22{{10$^{22}$}}

\def\feka{{Fe K$\alpha$}}
\def\xmogs{{\em EDOGs}}
\def\xmog{{\em EDOG}}
\def\3c{{3C 234}}
\def\l58{{$L_{5.8 \mu m}$}}

\begin{document}

\title{Revealing X-ray obscured quasars in SWIRE sources with extreme mid-IR/optical flux ratios}
\author{G. Lanzuisi\inst{1,2}, E. Piconcelli\inst{2},  F. Fiore\inst{2}, C. Feruglio\inst{3}, C. Vignali\inst{4,5}, M. Salvato\inst{6}, C. Gruppioni\inst{5}}
\titlerunning{Revealing obscured Quasars in the SWIRE survey}\authorrunning{G. Lanzuisi et al.}
\offprints{Giorgio Lanzuisi, \email{lanzuisi@oa-roma.inaf.it}}
\institute{Dipartimento di Fisica, Universit\`a  di Roma “La Sapienza”, P.le A. Moro 2, I--00185 Roma, Italy 
\and Osservatorio Astronomico di Roma (INAF), Via Frascati 33, I--00040
  Monteporzio Catone, Italy \and CEA, Irfu, Service d'Astrophysique, Centre de Saclay, 91191 Gif--sur--Yvette, France \and
Dipartimento di Astronomia, Universit\`a degli Studi di Bologna, Via Ranzani 1, I--40127 Bologna, Italy \and 
INAF--Osservatorio Astronomico di Bologna, Via Ranzani 1, I--40127 Bologna, Italy \and California Institute of Technology, MC 105--24, 1200 East California Boulevard, Pasadena, CA 91125, USA}

\date{}

\abstract{Recent works have suggested that selection criteria based on mid-IR  properties, i.e.
extreme colors and bright flux levels, can
be used to reveal a population of dust-enshrouded, extremely-luminous quasars at 
$z$ $\sim$1--2.
However, the X-ray spectral properties of these intriguing objects still remain 
largely unexplored.
We have performed  an X-ray study of a large sample of bright mid-IR 
(F$_{24\mu m}$ $>$ 1.3 mJy) galaxies showing an extreme MIR/Optical flux 
ratio (F$_{24\mu m}$/F$_R>$2000) in order to confirm the presence of a luminous active 
nucleus in these very red objects.
Sampling of a large area is required to pick up objects at the 
highest luminosities given their low surface density.
 Accordingly, we have applied our selection criteria to an area of $\sim$6 
deg$^2$ covered by \xmm/\chandra~observations within the $\sim$50 deg$^2$ SWIRE 
survey, resulting in a final sample of 44 objects.
The vast majority  of the source redshifts, both spectroscopic and photometric, are in the range 0.7 \simlt~$z$ \simlt~2.5.
The X-ray coverage of the sample is highly inhomogeneous (from 
snap-shot 5 ks \chandra~observations to medium-deep \xmm~exposures of 70 ks) 
and, consequently, a sizable fraction of them  ($\approx$43\%) 
remains undetected in the 0.5-10 keV band.
Using spectral or hardness information we were able to estimate the value of  
the absorbing column density in 23 sources. 95\% of 
them are consistent with being obscured by neutral gas with an intrinsic column 
density of \nh$\geq$ \e22~\cm2. Remarkably, we also find that $\sim$55\% of these 
sources can be classified as Type 2 quasars on the basis of their absorption 
properties and X-ray luminosity.
Moreover, most of the \xr~undetected sources show extreme mid-IR colors,
consistent with being luminous AGN-powered objects, suggesting they might host
heavily obscured (possibly Compton-thick) quasars in X-rays.
This demonstrates that our selection criteria applied to a wide area survey is very efficient 
in finding a large number of Type 2 quasars at $z$ \simgt~1. The existence of this class 
of very powerful, obscured quasars at high $z$ could have important 
implications in the context of the formation and cosmological evolution of 
accreting supermassive black holes and their host galaxies.}

\maketitle 

\section{Introduction}

The similarity in the anti-hierarchical
evolution of both star formation rate and AGN activity at $z<2$ (La
Franca et al. 2005; Arnouts et al. 2007)  reveals that the  assembly
and evolution of galaxies and those of the super massive black holes (SMBHs) at their centers are
intimately connected (e.g. Page et al. 2004; Alexander et al. 2005; Hopkins et al. 2006; Best et al. 2005; Croton et al. 2006).  
However, the census of accreting SMBHs through different cosmic epochs is still
highly incomplete and represents one of the major challenges in modern
extragalactic astronomy.
A clear-cut evidence for the existence of a large number of missing AGNs arises by the fact 
that the sources
detected in the recent \chandra/\xmm~surveys (e.g. Brandt \& Hasinger 2005) 
are able to resolve only  the 50--60\% of the cosmic X-ray background (CXB) in the 5-10 keV band (Worsley et al. 2005),
and their integrated radiation significantly fails to account for the CXB emission peak at $\sim$ 30 keV (Gilli, Comastri, Hasinger 2007; Comastri 2004). A population of heavily obscured AGNs showing a spectral shape consistent with the unresolved CXB component is usually invoked  to fill this gap.

Intriguingly, as stressed by Fiore et al. (2003), about 20-30\% of
the sources from \chandra/\xmm~surveys are optically-faint
(i.e. $R\simgt$25), and with a very high X-ray--to--optical flux ratio
(F$_{2-10}$/F$_R>$10). Optical spectroscopy of these extreme F$_{2-10}$/F$_R$ objects 
is therefore extremely challenging even with the largest telescopes.
Results from deep exposures and accurate multiwavelength studies
indicate that the vast majority of these sources could be obscured quasars at
high-$z$ (i.e. $z>$1; e.g. Mainieri et al. 2005;  Maiolino et
al. 2006). We have therefore started to uncover the long-sought
missing AGN population of the  X-ray absorbed, i.e. Type 2, quasars (QSO2s hereafter), that
is a key ingredient for both the CXB synthesis models as well as the
prediction of the SMBH/galaxy formation and evolution models
(e.g. Silk \& Rees 1998; Fabian 1999; Granato et al. 2004; Hopkins et
al. 2006). In particular, it has been hypothesized that before shining
in the optical as bright point-like blue objects, quasars undergo an
early dust enshrouded phase being associated with the rapid
SMBH growth. This obscured phase is likely
triggered by multiple galaxy encounters and mergers that favor the
infall and the concentration in the nuclear region of large reservoirs
of gas feeding the SMBH as well as the assembly of the massive
spheroids at high $z$. The link between the growth of the SMBHs and
their host galaxies is believed to be strong as witnessed by the tight
correlations between SMBH masses and bulge luminosity, mass and
velocity dispersion (Merritt \& Ferrarese 2001). A full understanding
of AGN feedback processes is therefore a key question to be addressed
in the context of  galaxy evolution (e.g. Di Matteo, Springel \&
Hernquist 2005; Alexander et al. 2005; Menci et al. 2008; Somerville et al. 2008).

In the latest years there have been many efforts aimed at unveiling
the missing population of heavily obscured quasars at high $z$
using alternative approaches with respect to the pure X-ray
selection.
Multiwavelength selection criteria based on the IR band seem to be very promising,
since absorbed AGN optical/UV/X-ray light  should be isotropically
re-emitted by the obscuring material at these frequencies.
In particular, multiple independent lines of evidence point towards selection
criteria using mid-IR (3--30 $\mu$m; MIR hereafter) photometry or combination of MIR and multiwavelenght data,
to efficiently collect large samples
of optically-faint quasars candidates at $z$\simgt~1 (e.g. Houck et
al. 2005; Yan et al. 2005; Weedman et al. 2006; Polletta et al. 2006;
Alonso-Herrero et al. 2006; Daddi et al. 2007; Hickox et al. 2007 Donley et al. 2008 and references therein). 
Combining MIR and radio data, Martinez-Sansigre et al. (2005) claimed the
discovery of a distant, optically-faint QSO2 population whose number
density is comparable with that observed for classical optically
bright quasars.  This might explain the paucity of obscured objects in
the high luminosity range of the X-ray selected AGN population that
has been reported in many works (Ueda et al. 2003; Steffen et
al. 2003; La Franca et al. 2005; but see also Dwelly \& Page 2006).
Furthermore, recent studies have provided strong evidence that a large fraction of the dust-obscured galaxies (DOGs hereafter) with extreme F$_{24\mu m}$/F$_R$ ratios (\simgt~1000) and bright at 24 $\mu$m (i.e. $F_{24\mu m}$ \simgt~0.5-1 mJy),  harbor
a high-$z$  obscured active nucleus (e.g. Dey et al. 2008; Polletta et al. 2008a; Georgantopoluos et al. 2008; Fiore et al. 2008a, 2008b).

However, an accurate determination of the
X-ray column density of the absorber in these sources is hampered by
the faintness of these sources in the X-ray band
(i.e. Polletta et al. 2006; Martinez-Sansigre et al. 2007). Therefore, most of the X-ray
information has been derived by the stacking of \chandra/\xmm~data or,
for small number of sources  detected with only a handful of counts,
by the hardness-ratio analysis.

Evidence for X-ray spectroscopically-confirmed heavily-absorbed QSO2s at cosmological distance still remains ambiguous  (i.e. Norman et al. 2002; Alexander et al. 2008; Erlund et al. 2008). Indeed, the Compton-thick (i.e. with  \nh\simgt$\sigma^{-1}_t$ $\approx$1.6 $\times$
10$^{24}$ \cm2) nature of the best Compton-thick quasar candidate,
IRAS 09104$+$4109, has been recently questioned (and not confirmed) by Piconcelli et al. (2007a) using \xmm~data. 
One of the reasons of this difficulty is that building up large and complete samples of bright (i.e. with a 2-10 keV flux $F_{2-10}$\simgt10$^{-14}$ \cgs) QSO2s by homogeneous selection criteria is complex and time-consuming,
because of the large area that must be covered with deep multi-band observations. This is necessary due to the low space density of the AGNs belonging to the bright end of the X-ray luminosity function.  
The Spitzer Wide-area InfraRed Extragalactic (SWIRE) survey (Lonsdale et al. 2003) has imaged nearly 50 square degrees in 6 fields: ELAIS-S1, ELAIS-N1, ELAIS-N2, Lockman Hole, XMM--LSS and CDFS with medium-deep MIPS and IRAC photometry, providing a unique opportunity to build up a large sample of luminous QSO2s. 

In this paper we report results on an X-ray spectroscopic study of a sample of SWIRE 
sources selected on the basis of their extreme    
mid-IR/optical flux ratios (F$_{24\mu m}$/F$_R>$2000) and  24 $\mu$m flux density (i.e. $F_{24\mu m}$ $>$1.3  mJy). 
These selection cuts have been applied in order to preferentially select sources harboring a luminous, dust-enshrouded AGN (e.g. Polletta et al. 2008a; Dey et al. 2008). Our aim is to confirm the presence of the active nucleus in these extreme DOGs and to directly measure the column density of the absorber via the characterization of their X-ray spectral properties.

The optical magnitudes used throughout the paper are Vega magnitudes.
The cosmological parameters used are $H_0$=70 km s$^{-1}$ Mpc$^{-1}$, $\Omega_\Lambda$ = 0.7 and $q_0$ = 0.

\begin{table*}
\caption{Optical-IR parameters and redshift of the sample.}
\label{tab:oir}
\begin{center}
\begin{tabular}{ccccccccccccccc}
\hline\hline\\
\multicolumn{1}{c} {ID}&
\multicolumn{1}{c} {Name}&
\multicolumn{1}{c} {R}&
\multicolumn{1}{c} {F$_{3.6\mu m}$}&
\multicolumn{1}{c} {F$_{4.5\mu m}$}&
\multicolumn{1}{c} {F$_{5.8\mu m}$}&
\multicolumn{1}{c} {F$_{8\mu m}$}&
\multicolumn{1}{c} {F$_{24\mu m}$}&
\multicolumn{1}{c} {Log(F$_{24\mu m}$/F$_R$)}&
\multicolumn{1}{c} {Log($\lambda L_{5.8\mu m}$)}&
\multicolumn{1}{c} {z}&
\multicolumn{1}{c} {z-type} \\
(1) & (2) & (3) & (4) & (5) & (6) & (7) & (8) & (9) & (10) & (11) & (12) \\\hline\\ 
1 &  J021559.46-045517.5  & 25.65     & 148.1 & 248.8  & 389.1 & 636.6 & 2364 & 4,2    & 45.14 &  1.007                  & SP $^a$   \\            
2 &  J021625.91-050425.1  & 24.39     & 14.31 & 34.45  & 106.5 & 400.0 & 2174 & 3.7    & 45.96 &  2.236                  & PH $^h$     \\              
3 &  J021656.59-051005.3  & 24.31     & 83.82 & 146.1  & 249.2 & 490.0 & 2083 & 3.7    & 45.12 &  1.080                  & PH $^h$    \\              
4 &  J021734.65-051718.2  & 26.36     & 59.84 & 111.0  & 222.1 & 419.7 & 1322 & 4.3    & 45.64 &  1.938                  & PH $^h$     \\              
5 &  J021743.09-044921.6  & 24.20     & 146   & 270.2  & 496.0 & 871.9 & 1508 & 3.5    & 45.13 &  1.042                  & PH $^h$     \\              
6 &  J021749.00-052306.9  & 22.36     & 375.6 & 714.2  &  1271 &  2317 & 8105 & 3.5    & 45.67 &  0.914                  & SP $^a$  \\              
7 &  J021754.45-043016.0  & 23.34     & 132.5 & 243.9  & 509.9 &  1055 & 2922 & 3.4    & 45.91 &  1.780                  & PH $^h$     \\              
8 &  J021912.66-043308.5  & 23.80     & 27.39 & 21.36  & $-$     & $-$ & 1471 & 3.3    & 44.42 &  0.754                  & PH $^h$     \\              
9 &  J021924.57-045300.0  & 25.26     & 30.38 & 42.14  & 67.16 & $-$   & 1473 & 3.9    & 45.62 &  1.979                  & PH $^h$     \\              
10 & J021935.19-050528.4  & 22.70     & 159.8 & 142.5  & 151.1 & 328.6 & 4095 & 3.3    & 44.76 &  0.826                  & SP $^b$   \\              
11 & J022003.58-045145.6  & 23.07     &   123 & 215.7  & 380.7 & 667.7 & 3374 & 3.4    & 45.28 &  1.080                  & PH $^h$     \\              
12 & J022003.95-045220.4  & 24.22     &  48.1 & 77.87  & 177.6 & 436.7 & 2831 & 3.7    & 45.50 &  1.443                  & PH $^h$     \\              
13 & J160623.44+542301.9  & 24.59     & 35.85 & 41.66  & 43.79 & 86.93 & 1546 & 3.6    & 45.05 &  1.466                  &PH$^h$  \\       
14 & J160650.71+542510.1  & $>$24.30  & 290.2 & 258.2  & 193.1 & 818.8 & 1768 & $>$3.5 & 43.66 &  0.22                   &SP$^c$  \\       
15 & J160700.62+542416.8  & 26.02     & 18.04 & 37.57  & 95.66 & 315.3 & 1937 & 4.2    & 44.35 &  0.63$_{-0.20}^{+0.20}$ &PH$^g$  \\       
16 & J160813.86+541942.4  & 25.09     & 63.38 & 130.0  & 270.4 & 531.2 & 2631 & 4.0    & 45.09 &  1                      &PH$^g$  \\     
17 & J160816.15+545351.1  & 24.12     &  64.9 & 127.7  & 333.1 & 680.6 & 2147 & 3.5    & 44.75 &  0.70$_{-0.15}^{+0.15}$ &PH$^g$  \\  
18 & J160913.02+540509.7  & 25.31     & 61.84 & 136.9  & 292.7 & 581.3 & 1319 & 3.8    & 45.02 &  1                      &PH$^g$  \\  
19 & J160913.28+542322.0  & 23.86     & 254.4 & 478.6  & 987.4 & 1966  & 4798 & 3.8    & 45.90 &  1.400                  &PH$^h$  \\  
20 & J160933.78+535720.2  & 25.57     & 50.39 & 69.80  & 60.08 & 74.27 & 1373 & 3.9    & 44.43 &  1                      &PH$^g$  \\ 
21 & J160950.98+535836.3  & 23.95     & 131.9 & 236.4  & 406.9 & 752.3 & 1692 & 3.3    & 45.26 &  1.138                  &PH$^h$  \\ 
22 & J161007.17+540628.5  & 23.34     & 38.47 & 90.59  & 271.9 & 737.3 & 3284 & 3.4    & 45.95 &  1.871                  &PH$^h$  \\ 
23 & J161017.70+535124.1  & $>$24.80  & 17.48 & 37.81  & 80.36 & 260.6 & 1572 & $>$3.7 & 44.37 &  0.69$_{-0.20}^{+0.20}$ &PH$^g$   \\ 
24 & J161017.93+542332.1  & 23.89     & 36    & 58.41  & 135.9 & 375.9 & 2273 & 3.5    & 44.42 &  0.63$_{-0.10}^{+0.10}$ &PH$^g$   \\ 
25 & J161037.63+540603.5  & 24.88     & 12.05 & 12.63  & $-$   & 50.06 & 1724 & 3.7    & 46.58 &  3.77$_{-0.77}^{+0.48}$ &PH$^g$   \\ 
26 & J161051.03+535330.5  & 25.39     & 13.26 & 20.86  & 57.56 & 197.4 & 1560 & 3.9    & 44.22 &  0.66$_{-0.15}^{+0.15}$ &PH$^g$   \\ 
27 & J161052.74+543740.4  & 23.53     & 70.77 & 66.55  & 72.03 & 136.1 & 2646 & 3.4    & 45.21 &  1.410                  &PH$^h$  \\ 
28 & J161102.22+550551.6  & 24.10     & 71.53 & 73.24  & 121.4 & 350.9 & 2661 & 3.6    & 44.97 &  0.995                  &PH$^h$   \\
29 & J161204.84+544149.2  & 25.07     & 33.74 & 46.92  & 42.97 & 54.68 & 1808 & 3.8    & 45.98 &  2.58$_{-0.30}^{+0.40}$ &PH$^g$    \\
30 & J161306.28+543959.7  & 24.91     &  23.5 & 54.15  & 170.7 & 548.9 & 2922 & 4.0    & 44.40 &  0.54$_{-0.10}^{+0.10}$ &PH$^g$  \\
31 & J163723.94+410525.7  & 23.80     & 77.52 & 65.69  & 52.22 & 133.9 & 1856 & 3.3    & 44.94 &  1.24                  &PH$^h$       \\       
32 & J104409.95+585224.7  & 23.56     & 64.27 & 150.4  & 390.9 &  1093 & 4134 & 3.6    & 46.41&   2.54                   &SP$^d$  \\ 
33 & J104528.29+591326.6  & 23.68     & 32.18 & 45.19  & 90.94 & 206.6 & 2462 & 3.4    & 45.98&   2.31                   &SP$^e$   \\
34 & J104837.82+572816.2  & 23.84     & 100.6 & 170.3  & 307.5 & 625.3 & 2355 & 3.5    & 45.53&   1.443                  &PH$^h$    \\ 
35 & J104847.14+572337.6  & 23.74     & 158.1 & 285.4  & 451.8 & 797.1 & 2590 & 3.5    & 45.42&   1.218                  &PH$^h$    \\ 
36 & J104954.96+584429.4  & 24.15     & 18.45 & 23.43  & 58.82 & 146.6 & 1378 & 3.3    & 44.27&   0.76$_{-0.10}^{+0.10}$ &PH$^g$   \\    
37 & J105604.84+574229.9  & 22.21     & 62.28 & 85.63  & 158.6 & 438.5 & 11280& 3.5    & 45.59&   1.249                  &PH$^h$    \\ 
38 & J003314.61-432300.3  & 24.13     &  10.5 & 16.69  & 57.68 & 142.5 & 1648 & 3.4    & 44.46 &  0.87$_{-0.20}^{+0.20}$  & PH$^g$\\       
39 & J003316.92-431706.3  & 24.30     & 142.2 & 295.1  & 581.9 &  1059 & 2197 & 3.6    & 44.88 &  0.689                   & SP$^f$\\        
40 & J003333.75-432326.9  & 24.41     & 48.02 & 73.56  & 161.1 & 225.4 & 1493 & 3.5    & 45.03 &  1.24$_{-0.20}^{+0.20}$  & PH$^g$\\       
41 & J003336.26-431731.7  & 24.65     & 18.66 & 34.14  & 82.01 & 259.0 & 2188 & 3.7    & 44.51 &  0.76$_{-0.10}^{+0.10}$  & PH$^g$\\       
42 & J003348.18-432822.3  & $>$26.63  & 16.17 & 40.72  & 98.36 & 346.5 & 1682 & $>$4.4 & 44.70 &  0.84$_{-0.05}^{+0.05}$  & PH$^g$\\       
43 & J003518.20-433414.6  & 25.19     & 12.06 & 16.96  & $-$   & 142.6 & 1745 & 3.9    & 45.38 &  1.69$_{-1.00}^{+0.40}$  & PH$^g$\\       
44 & J003641.47-432038.1  & 23.91     &   110 & 224.6  & 459.3 & 911.7 & 2746 & 3.6    & 45.25 &  1                       & PH$^g$\\        
 \hline                        
\end{tabular}\end{center}                 
Column: 
(1) ID number; 
(2) Name from SWIRE2 catalogue;
(3) Magnitude R; 
(4) 3.6 $\mu$m Flux density in $\mu$Jy;
(5) 4.5 $\mu$m Flux density in $\mu$Jy;
(6) 5.8 $\mu$m Flux density in $\mu$Jy;
(7) 8 $\mu$m Flux density in $\mu$Jy;
(8) 24 $\mu$m Flux density in $\mu$Jy;
(9) Logarithm of F$_{24\mu m}$/F$_R$ flux ratio; 
(10) Logarithm of 5.8 $\mu$m Luminosity in erg s$^{-1}$;  
(11) Redshift;
(12) Type of redshift: SP spectroscopic redshift [$^a$ this work (see section 2.3.1 for details); $^b$ Simpson et al. 2006; $^c$ from Sloan Digital Sky Survey Data Release 2; $^d$ Polletta et al. (2006); $^e$ Weedman et al. 2006a; $^f$ Feruglio et al. 2008], PH photometric redshift [$^g$ this work (see section 2.3 for details); $^h$ Rowan-Robinson et al. 2008; $^i$ fixed 
value of $z$].
\end{table*}
\section{The sample: selection criteria and datasets}
\label{sample}

\subsection{Optical and IR observations}

Infrared and optical photometric data are available from the public archive of the SWIRE survey (Lonsdale et al. 2003 and references therein).
The catalog contains medium-deep photometric data in MIPS 24 $\mu m$ band down to 400 $\mu$Jy, 5$\sigma$, and in the four IRAC channels from 3.6 to 8.0 $\mu m$  (4.1 $\mu$Jy at 3.6 $\mu m$, 5$\sigma$, Surace at al. 2004).
Optical photometry is also available in different bands for selected areas in each of the 5 SWIRE fields used in this work (ELAIS-S1, ELAIS-N1, ELAIS-N2, Lockman Hole, \xmm~SXDS). We did not consider the CDFS here since these data were already studied by Fiore et al. (2008a) using a F$_{24\mu m}$/F$_R$ selection technique similar to our approach.
ELAIS-N1 and N2 have optical photometric information in 5 bands (U'g'r'i'Z') from the Wide Field survey (WFS, McMahon et al. 2001). The
Lockman Hole sky region is covered with 4-band Ug'r'i' photometry from the SWIRE photometry programme. 
XMM-LSS has very deep ($R_{AB}<27.7$) 5 band BVRi'z' photometry in 1.12 deg$^{2}$ from the Subaru XMM Deep Survey (SXDS, e.g. Sekiguchi et al. 2005; Furusawa et al. 2008). 
Full details on the optical and IR photometry for the ELAIS-S1 field can be found in Feruglio et al. (2008).

 We firstly selected sources showing a 24 $\mu m$ flux density F(24$\mu m)>1000~ \mu$Jy and F$_{24\mu m}$/F$_R>1000$ in order to have an initial sample of DOGs (560 sources e.g. Fig. 1a.
Several sources in ELAIS-S1, ELAIS-N1, ELAIS-N2 and Lockman Hole do not show an R magnitude value in the catalogs mostly since
they are faint objects, i.e. with a magnitude above the detection limit.
For these sources we performed aperture photometry at the position of the MIPS source.
The photometry was performed using the IRAF \emph{apphot} tool, with an 1.2 arcsec aperture radius (consistent with the released optical aperture photometry), and a 10 arcsec wide annulus for background estimate.
For undetected sources we computed a 3$\sigma$ limiting magnitude using the median value of $\sigma_{rms}$ at the positions of the source and the relation
\begin{equation}
R_{lim}=-2.5\times log(\sqrt(A)\times 3 \times \sigma_{rms}) + zero point
\end{equation}
where A is the area corresponding to the aperture. 

\subsection{Selection criteria}

We selected all the sources with available X-ray data from the initial sample of DOGs (i.e. F(24$\mu m)>1000~ \mu$Jy and F$_{24\mu m}$/F$_R>1000$ sources). The portion of sky covered by \xmm~and/or \chandra~observations (for a total of $\sim6$ deg$^{2}$) significantly changes across the five SWIRE fields discussed here, varying from $\sim$100\% for the XMM-SXDS field to $\sim$5\% for the ELAIS-N2 field. Nonetheless, since such a selection is completely random and not based on specific target observations, it should not introduce any selection effect in the final sample.

\begin{figure*}
\begin{center}
\includegraphics[width=8cm,height=8cm]{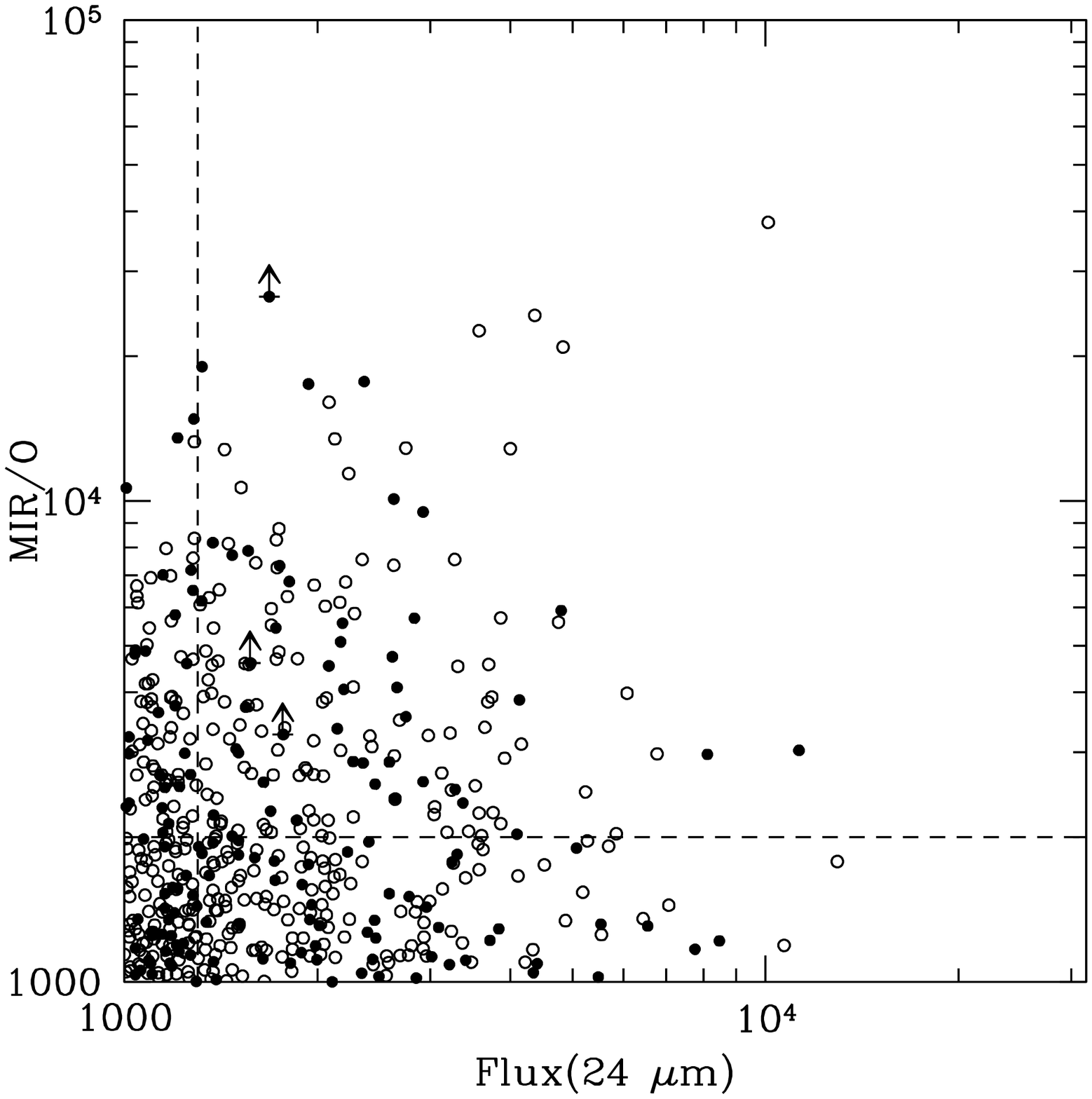}\hspace{1cm}\includegraphics[width=8cm,height=8cm,angle=0]{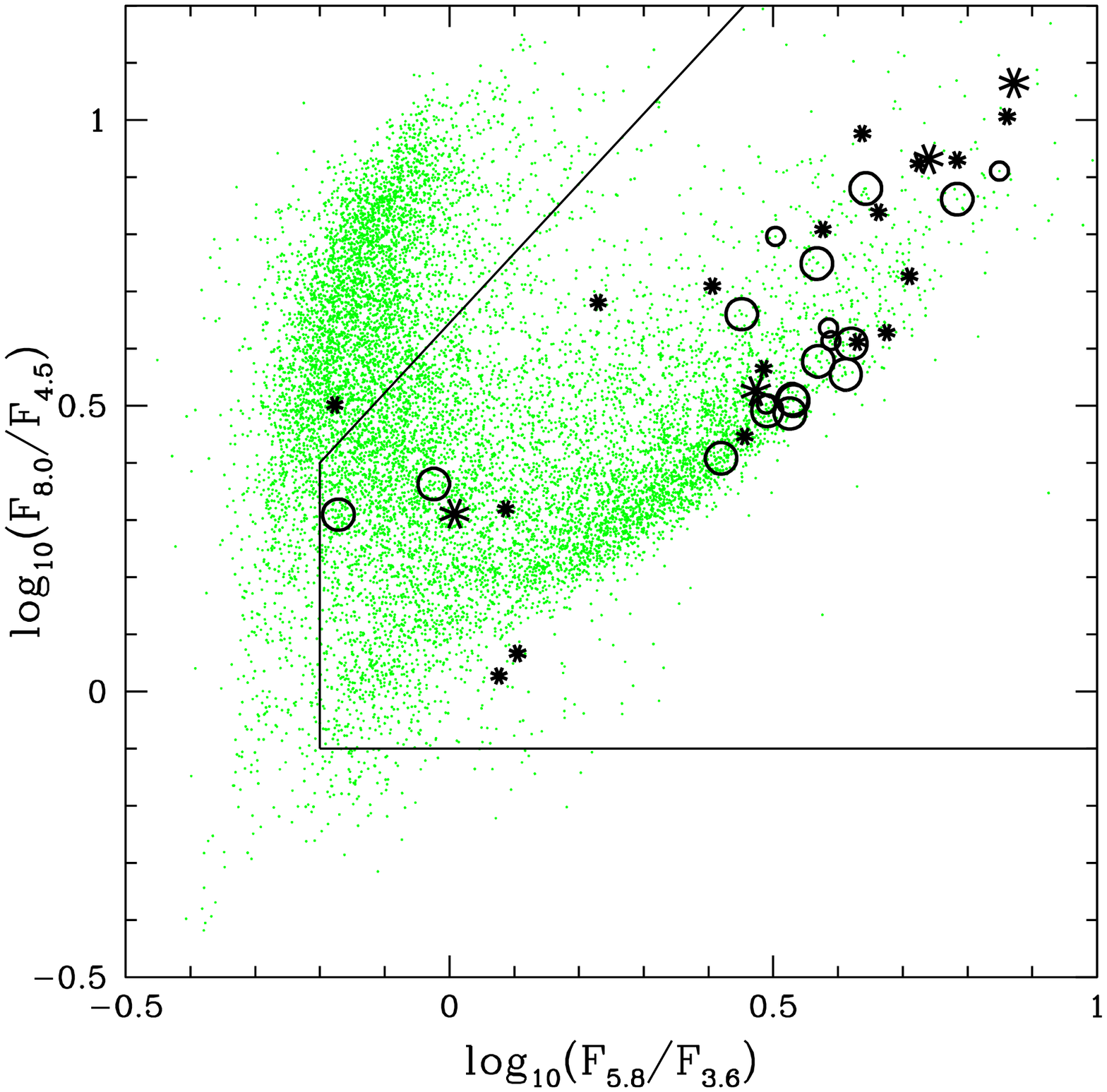}
\caption{{\it a) Left panel:} 24 $\mu$m flux density vs. F$_{24\mu m}$/F$_R$ in logarithmic scales for all the sources with F$_{24\mu m}$/F$_R$ $\geq1000$ and F$_{24\mu m}\geq1$ mJy in the 5 SWIRE fields considered in this work. Sources with (without) available X-ray coverage are represented with filled (open) circles. Dashed lines represent the selection criteria: F$_{24\mu m}$/F$_R$ $\geq$2000 and 24 $\mu$m flux density $\geq$ 1.3 mJy. {\it b) Right panel:} An IRAC color-color diagram of the SWIRE $F_{24\mu m} > $ 1 mJy sample (small points). Sources in our sample with (without) \nh~estimate are plotted with open circles (asterisks). Small (large) symbols represent sources with effective exposure time $\leq$ 15 ks ($\geq$ 15 ks). The solid line marks the region of IRAC colors typical of AGN (Lacy et al. 2004).}
\end{center}
\label{fig:lacy}
\end{figure*}

We started selecting sources with F$_{24\mu m}$/F$_R>$2000 at the highest F$_{24\mu m}$ flux density levels since bright 24$\mu m$ objects with large F$_{24\mu m}$/F$_R$ are predominantly MIR power-law (i.e. AGN-dominated) sources (e.g. Dey et al. 2008; Donley et al. 2008).
We then gradually decreased the F$_{24\mu m}$ flux density threshold until a statistically significant sample of objects was collected.
In fact, fainter sources (F$_{24\mu m}\sim$ 0.1 mJy) selected with the same extreme F$_{24\mu m}$/F$_R$ colors, turn out to be dominated by star-formation activity both in the \xr~and MIR bands (Donley et al. 2008; Pope et al. 2008a).
Accordingly we selected all the sources showing a F$_{24\mu m}$/F$_R\geq$2000 and a 24 $\mu$m flux density F(24$\mu m$) $\geq$ 1.3 mJy (see Fig. 1a) obtaining in this way a sample of 50 objects, referred as Extreme DOGs (i.e. \xmogs) hereafter.
Since in four cases the same optical object has a couple of IRAC-MIPS counterparts, we retained as the most likely counterpart the MIR source located at the smallest distance from the optical position.

In this way we got a sample of 46 \xmogs~within the 6 deg$^2$ of SWIRE survey covered by \xr~observations. This implies a number density of 7.7 deg$^{-2}$ that is in very good agreement with that obtained applying the same selection criteria to the sample of DOGs in the $\approx$8.6 deg$^{2}$ of the NOAO Deep Wide-Field Survey $Bo\ddot{o}tes$ field analyzed in Dey et al. 2008 (i.e. 7.6 sources deg$^{-2}$). The fraction of sources with F$_{24\mu m}$/F$_R\geq$2000 among all the sources  with F(24$\mu m$) $\geq$ 1.3 mJy is $\sim$10\%.

Two sources (namely SWIRE2 J160734.22+544217.3 and SWIRE2 J160945.43+534944.6) fall in a CCD gap in the X-ray image and therefore were excluded from the sample since there is no useful X-ray information for them. 
The final sample therefore consists of 44 sources and is listed in Table 1.
Fig. 1b shows the MIR color-color diagram (e.g. Lacy et al. 2004) for the sources with F$_{24\mu m}>$1 mJy in the 5 SWIRE fields considered in this work (SWIRE F$_{24\mu m}>$1 sample hereafter) obtained using the flux densities measured in the four {\it Spitzer-}IRAC channels\footnote{Source Nos. 8, 9, 25 and 43 are undetected in one (or two) IRAC bands and, therefore,  they are not plotted in Fig. 1b.}.
Two distinct populations can be readily separated in this plot. One has blue colors in F$_{5.8}$/F$_{3.6}$ and red colors in F$_{8.0}$/F$_{4.5}$, most likely consisting of a mixture of starlight-dominated galaxies and 
 star-forming galaxies with MIR 5--12$\mu$m spectra dominated by PAH emission features at low-redshift ($z$\simlt~0.2--0.4). 
The other population has redder colors and can be mainly identified with  galaxies with a MIR power-law spectrum that is a typical signature of an AGN-powered emission.
The thick lines mark the region of the diagram typically occupied by AGNs (Lacy et al. 2004; Alonso-Herrero et al. 2006; Polletta et al. 2006; Lacy et al. 2007; Barmby et al. 2008; see also Donley et al. (2008) for a detailed discussion about the level of contamination by star-forming galaxies in this region).
It is worth noting  that all but one  sources in our sample lie within the AGN region, and most of them show extreme red colors i.e. log(F$_{5.8}$/F$_{3.6}$)\simgt~0.4 and log(F$_{8.0}$/F$_{4.5}$)\simgt~0.4 similarly to the luminous obscured quasars at $z$=1.3--3 in the Polletta et al. (2008a) sample. The only source falling outside the box has the lowest spectroscopic redshift in the sample, i.e. source No. 14 at $z$=0.22, and therefore it is likely that the contribution from the host galaxy stellar light can explain the low value of the F$_{5.8}$/F$_{3.6}$ ratio. The range of MIR colors observed for our \xmogs~can be likely ascribed to the relative strength of the  contribution of star formation to the total MIR luminosity (e.g. Dey et al. 2008; Sajina, Lacy \& Scott 2005).
Interestingly, Donley et al. (2007)  and Cardamone et al. (2008) found that the most luminous X-ray sources with \lum\simgt~5 $\times$ 10$^{44}$ \ergs~in their large samples of MIR sources fall in this box of the IRAC color-color diagram.

\subsection{Redshifts}
\label{sec:z}

\begin{figure*}
\begin{center}
\includegraphics[width=6cm,height=8cm,angle=-90]{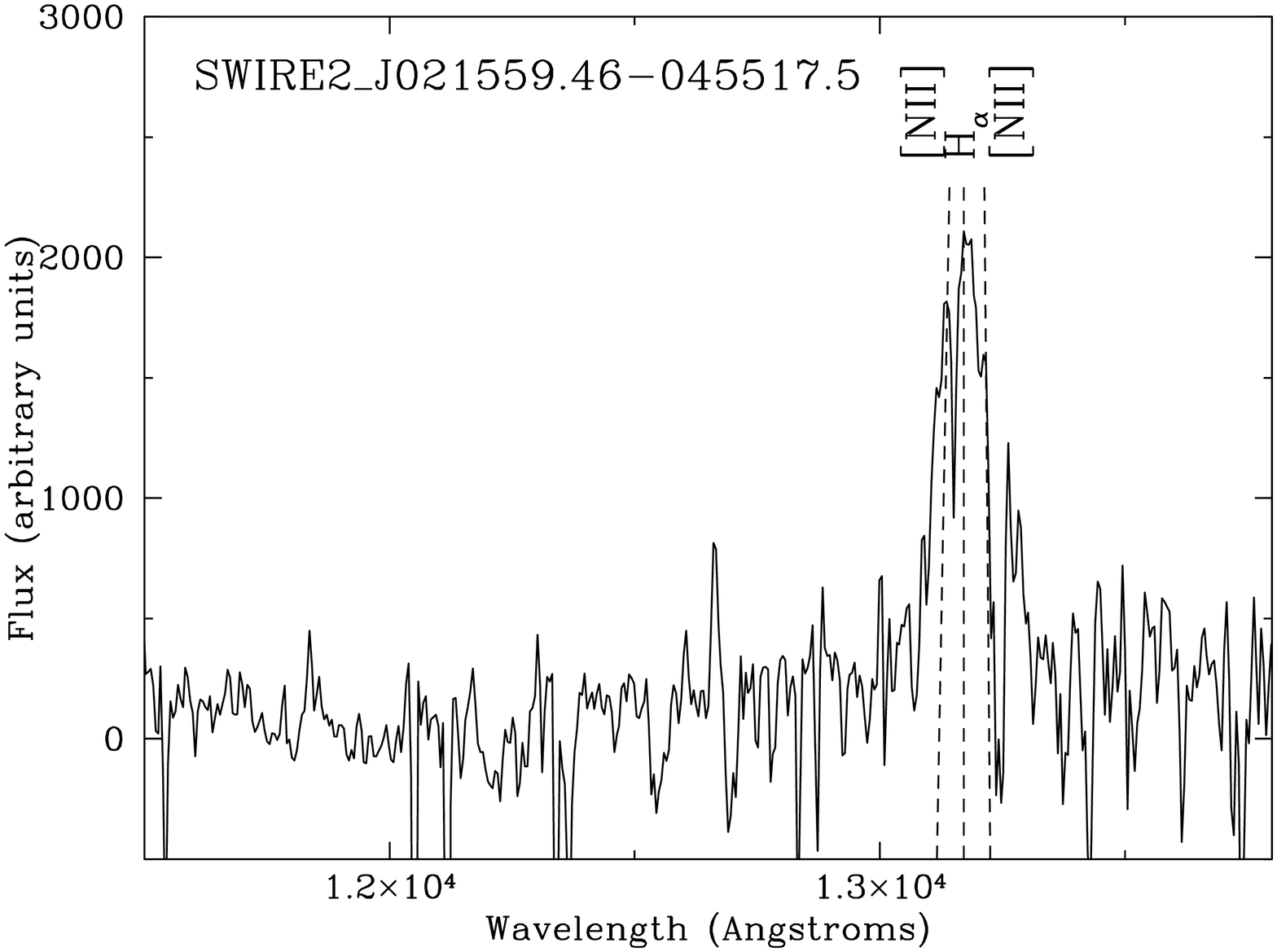}\hspace{1cm}\includegraphics[width=6cm,height=8cm,angle=-90]{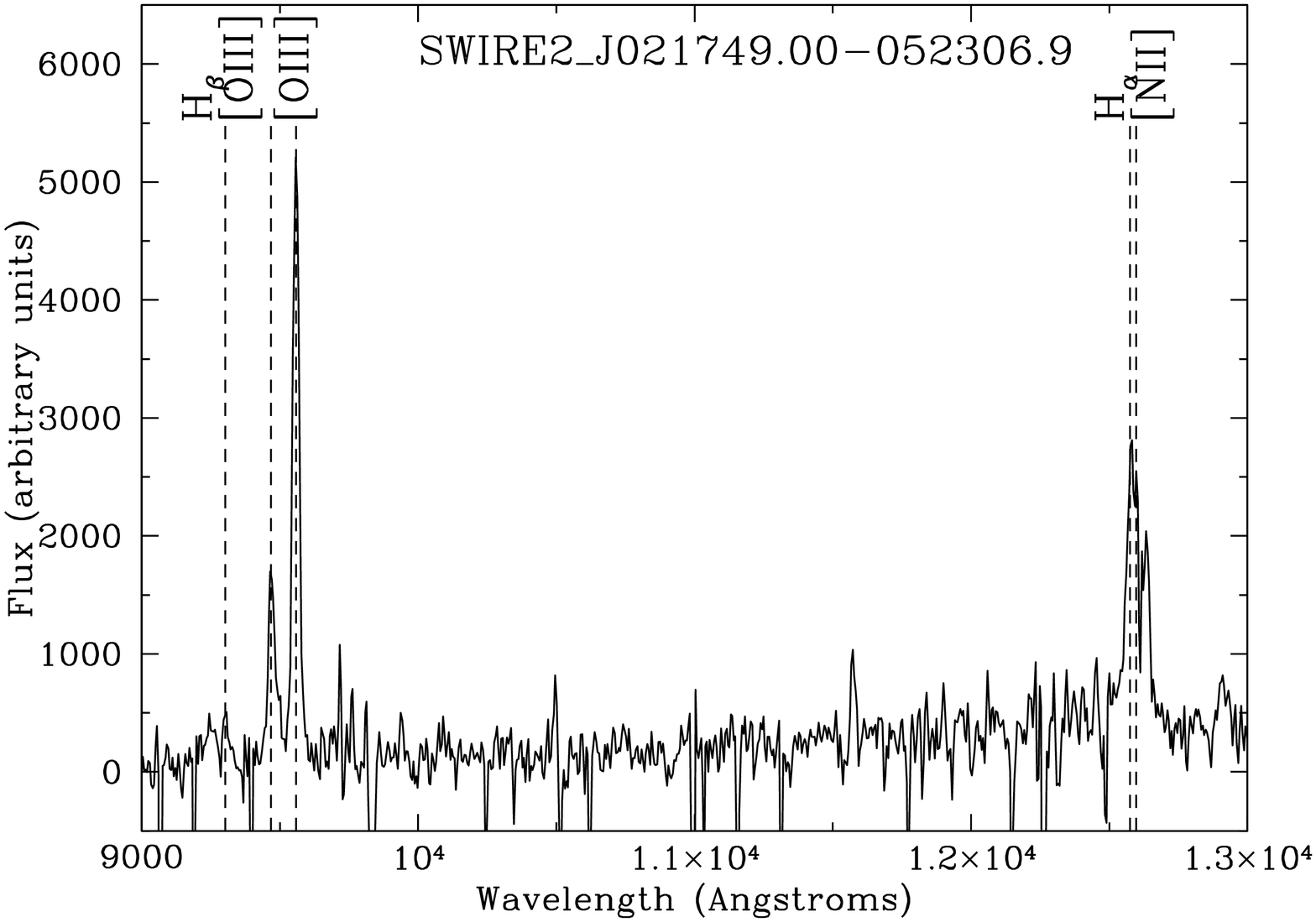}
\caption{{\it a) Left panel:} Near IR spectrum of source No. 1 at z=1.007. {\it b) Right panel:} Near IR spectrum of source No. 6 at z=0.914. Both spectra were obtained by MOIRCS/SUBARU observations.}
\end{center}
\end{figure*}

Spectroscopic redshifts are available for 7 objects in the sample (0.22 $\leq$ z$_{\rm sp}$ $\leq$ 2.54; e.g. Table 1).
In particular, two of them are obtained from new observations performed with the Subaru telescope (see Section 2.3.1 for details of data reduction and analysis). 

For the remaining 37 sources we were able to derive photometric redshifts thanks to the available optical and IR photometric data. We used the code described in Fontana et al. (2000) and Fontana (2001).
Photometric redshifts were calculated using a $\chi^2$ minimization technique on spectral libraries which include  starburst, passive galaxy  and AGN semi-empirical template spectra from Polletta et al. (2007), Fiore et al. (2008a) and Pozzi at al. (2007).
All the available photometric data points, from the optical to the MIPS 24 $\mu m$ band, were used in the computation (e.g. Feruglio et al. 2008).
The photo-$z$ are reported in Table 1 with errors corresponding to a confidence level of 1.6 $\sigma$ and their distribution spans in the range 0.54 $\leq z_{ph} \leq$ 3.77. 

In four cases the $\chi^2$ minimization technique was not able to properly constrain the $z$ value. For these sources (namely Nos. 16, 18, 20 and 44) we decided to assume a fixed value of $z$=1 which is close to the mean value of the $z$ distribution in our sample ($\langle z\rangle \sim$1.2). It is worth noting that this is a conservative assumption since the typical value of z found for large samples of DOGs is $\sim2$ (e.g. Polletta et al. 2008a, Dey et al. 2008 and references therein).

For 19 of 37 sources photometric redshifts from the SWIRE Photometric Redshift Catalogue published in Rowan-Robinson et al. (2008) are available\footnote{ For 15 sources we were able to compare the photo-$z$ from the SWIRE Photometric Redshift Catalogue with those computed with our method. We found that for $\sim$50\%($\sim$85\%) of the sources, the photo-$z$ values are consistent within $15\%$ ($30\%$).}.
Their approach is based on galaxy and quasar templates applied to data at 0.36$-$4.5 $\mu$m, and on a set of 4 IR emission templates fitted to IR excess data at 3.6$-$170 $\mu$m.
The photo-$z$ distribution from the SWIRE Photometric Redshift Catalogue spans in the range 0.754 $\leq z_{ph} \leq$ 2.236.

\begin{figure*}
\begin{center}
\includegraphics[width=8cm,height=8cm]{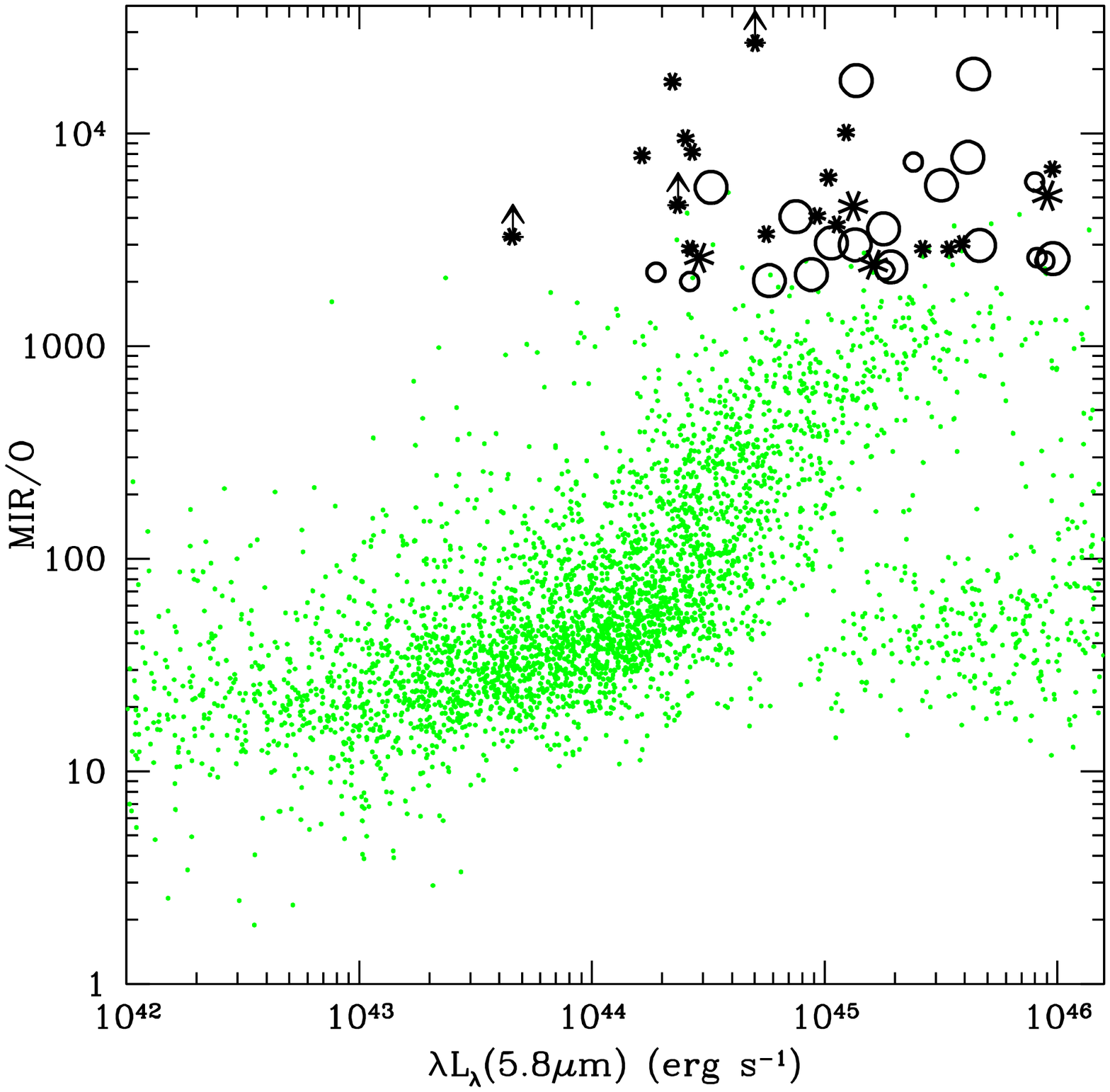}\hspace{1cm}\includegraphics[width=8cm,height=8cm]{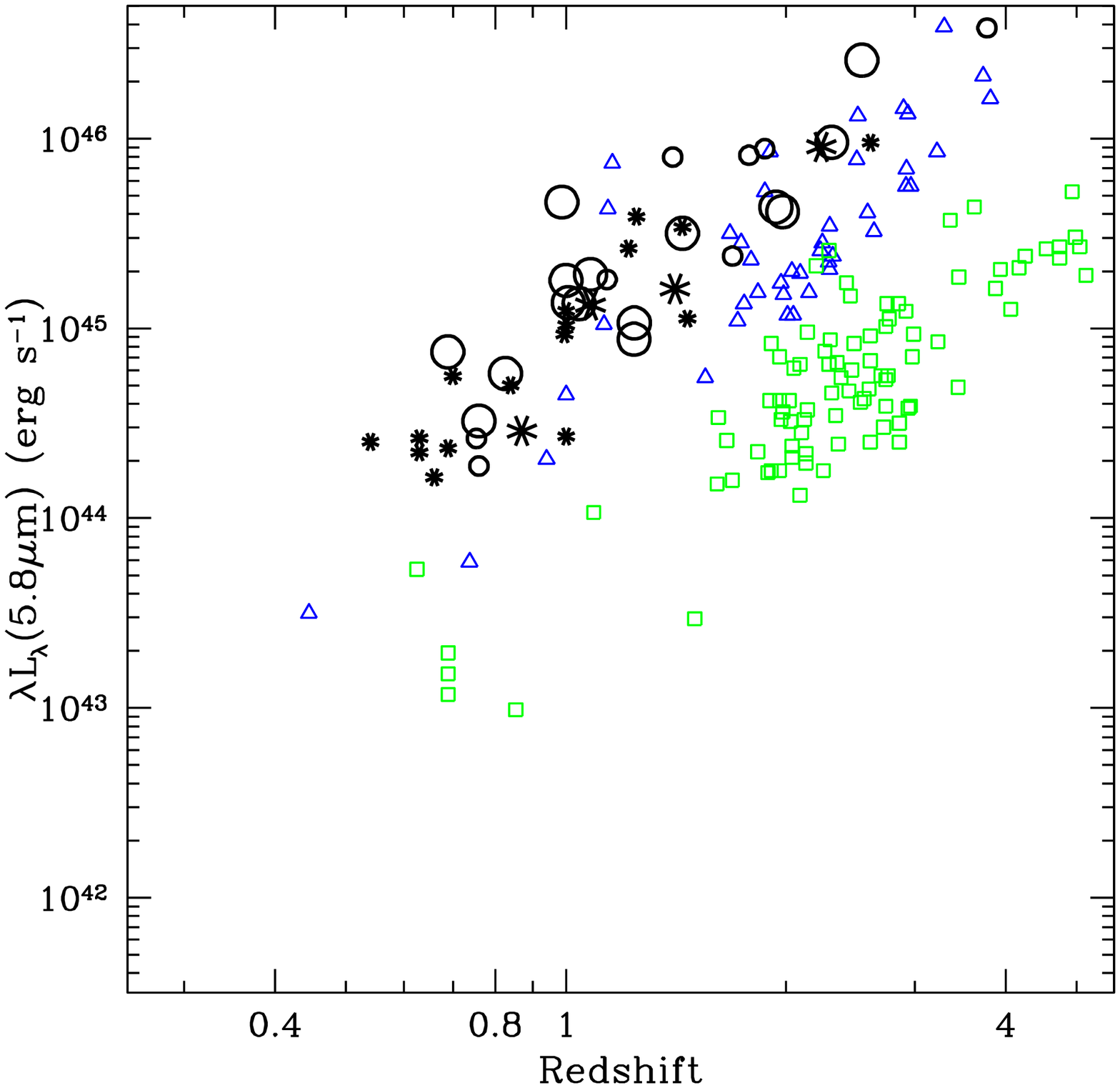}
\caption{{\it a) Left panel:} $\lambda L_{5.8 \mu m}$ vs. F$_{24\mu m}$/F$_R$ ratio for the SWIRE $F_{24\mu m} > $ 1 mJy sample (small points). Sources in our sample with (without) \nh~estimate are plotted with open circles (asterisks). Small (large) symbols represent sources with effective exposure time $\leq$ 15 ks ($\geq$ 15 ks). {\it b) Right panel:} The $\lambda L_{5.8 \mu m}$ -- redshift plane for sources in our sample (symbol as in the left panel), COSMOS sources (open triangles) and CDFS sources (open squares) with F$_{24\mu m}$/F$_R>$2000 from Fiore et al. (2008a, b).}
\end{center}
\end{figure*}

\subsubsection{Spectroscopic redshifts}

We obtained the spectroscopic redshifts for two sources (namely source Nos. 1 and 6) via near--IR spectroscopy.
These observations were performed  with MOIRCS (Multi-Object  
InfraRed Camera and Spectrograph) mounted on the Subaru telescope  
(Hawaii) on 2007 Dec 22, as backup program (program No. S07B0087N, P.I.  
M. Salvato).
We  used  the zJ500 grism  (optimized for the observation of 0.9-1.78  
$\mu$m region) with a R500 filter blocking the contamination of the  
second-order spectra at $\lambda>$1.5 $\mu$m.
Each observation consists of 2 frames  dithered  along the slit by 5 arcsec.

After standard reduction  the spectra have been subtracted by each other   
in order  to improve sky subtraction and flat-fielding and then  
combined.
SWIRE2 J021559.46-045517.5 has been observed for a total of 1200 sec and SWIRE2 J021749.00-052306.9 for a total of  
1800 sec.
Sky lines have been used for the wavelength calibration.
Results can be summarized as follows:
\begin{itemize}
\item {\it SWIRE2 J021559.46-045517.5 (No. 1):}
We obtained a redshift of z=1.007 on the basis of the identification of H$_{\alpha}\lambda6563$ (FWHM $<$ 1100 km s$^{-1}$) [NII]$\lambda6548$ and [NII]$\lambda6583$ narrow emission lines.  (see Fig. 2a).
\item {\it SWIRE2 J021749.00-052306.9 (No. 6):} 
We obtained a redshift of z=0.914 on the basis of the identification of strong [OIII]$\lambda4959,\lambda5007$, H$_{\alpha}\lambda6563$ and [NII]$\lambda6583$ narrow emission lines (see Fig. 2b). This value does not match with the tentative redshift of z=0.987 proposed by Lacy et al. (2007), using very low S/N optical spectral data. The FWHM of the H$_{\alpha}$ line is FWHM$<$1000 km s$^{-1}$. This support a narrow line quasar classification for this \xmog.
\end{itemize}

\begin{table*}
\caption{X-ray coverage and 0.5-10 keV band flux of the sample.}
\label{tab:xray}
\begin{center}
\begin{tabular}{ccccccc}
\hline\hline\\
\multicolumn{1}{c} {ID}&
\multicolumn{1}{c} {Exposure}&
\multicolumn{1}{c} {Telescope}&
\multicolumn{1}{c} {CR$_{0.5-10}$}&
\multicolumn{1}{c} {Flux$_{0.5-10}$}&
\multicolumn{1}{c} {Log(F$_{2-10}$/F$_R$)}\\
 (1) & (2) &(3) & (4) & (5) & (6) \\
\hline\\ 
1 & 18.8  & {\em XMM--N}  & 25.6$\pm$4.3 &  18.2$\pm$3.1         &50.1  $\pm$   8.4 &\\
2 & 23.2  & {\em XMM--N}  & 2.1$\pm$0.9  &  1.8$\pm$0.8          &$<$2.0            &\\
3 & 29.3  & {\em XMM--N}  & $<$2.4       &  $<$2.1               &$<$2.3            & \\
4 & 38.0  & {\em XMM--N}  & 3.9$\pm$1.2  &  9.4$\pm$5.1          &79.4  $\pm$ 24.4  & \\
5 & 31.8  & {\em XMM--N}  & 3.5$\pm$1.8  &  2.4$\pm$1.1          &0.4   $\pm$ 0.2   & \\
6 & 32.1  & {\em XMM--N}  & 86.9$\pm$5.1 &  111$\pm$5.5          &131.8 $\pm$ 7.7   & \\ 
7 & 6.8   & {\em XMM--N}  & 11.5$\pm$2.0 &  15.6$\pm2.7$         &8.1   $\pm$ 3.4   &\\
8 & 9.0   & {\em XMM--N}  & 3.2$\pm$1.6  &  3.7$\pm1.6$          &3.0   $\pm$  1.5  &\\
9 & 21.7  & {\em XMM--N}  & 4.7$\pm$1.9  &  6.0$\pm2.3$          &18.2  $\pm$ 9.7   &\\
10 & 23.7 & {\em XMM--N}  & 2.1$\pm$1.5  &  2.6$\pm$1.6          &$<$ 0.7           & \\
11 & 17.0 & {\em XMM--N}  & 8.5$\pm$3.3  &  6.8$\pm2.3$          &2.6   $\pm$ 1.0   & \\
12 & 17.6 & {\em XMM--N}  & 17.6$\pm$3.0 &  29.1$\pm9.0$         &37.2  $\pm$ 14.7  & \\
13 & 4.7  & \chandra~     & $<$13.6      &  $<$16.3              &$<$22.9           &      \\
14 & 4.4  & \chandra~     & $<$23.7      &  $<$41.9              &$<$53.24          &      \\
15 & 4.4  & \chandra~     & $<$14.2      &  $<$30.5              &$<$199.5          &      \\
16 & 4.4  & \chandra~     & $<$14.3      &  $<$25.3              &$<$67.6           &      \\
17 & 4.1  & \chandra~     & $<$20.6      &  $<$42.5              &$<$47.9           &      \\
18 & 4.5  & \chandra~     & $<$19.0      &  $<$33.6              &$<$107.2          &      \\
19 & 4.6  & \chandra~     & 10.3$\pm$4.9 &  15.0$\pm8.7$         &11.7  $\pm$ 5.6   &     \\
20 & 4.7  & \chandra~     & $<$23.9      &  $<$42.3              &$<$173.8          &  \\
21 & 4.5  & \chandra~     & 23.4$\pm$7.3 &   30.1$\pm6.4$        &25.1  $\pm$ 7.8   &   \\
22 & 7.0  & {\em XMM--N}  & 12.7$\pm$4.7 &   13.4$\pm5.2$        &6.5   $\pm$ 2.4   &  \\
23 & 4.4  & \chandra~     & $<$14.2      &  $<$29.5              &$<$47.59          & \\
24 & 4.8  & \chandra~     & $<$17.9      &  $<$37.8              &$<$34.7           & \\
25 & 4.6  & \chandra~     & 15.3$\pm$5.2 &  22.1$\pm5.5$         &44.7  $\pm$ 15.2  &  \\
26 & 4.4  & \chandra~     & $<$23.6      &  $<$49.8              &$<$182.0          & \\
27 & 68.4 & \chandra~     & $<$2.1       &  $<$3.7               &$<$2.3            &  \\
28 & 4.4  & \chandra~     & $<$13.7      &  $<$23.6              &$<$24.5           & \\
29 & 4.7  & \chandra~     & $<$22.0      &  $<$25.2              &$<$53.7           & \\
30 & 4.5  & \chandra~     & $<$20.0      &  $<$45.3              &$<$107.2          & \\
31 & 66.9 & \chandra~     &1.4$\pm$0.5   &   2.0$\pm0.8$         &1.5   $\pm$ 0.6   & \\ 
32 & 67.8 & \chandra~     & 2.3$\pm$0.6  &  3.6$\pm0.8$          &2.2   $\pm$  0.6  & \\
33 & 66.6 & \chandra~     & 2.6$\pm$0.7  &  4.4$\pm0.8$          &3.0   $\pm$ 0.8   & \\
34 & 4.7  & \chandra~     & 6.3$\pm$3.7  &  10.4$\pm6.1$         &$<$8.5            & \\
35 & 4.6  & \chandra~     & $<$14.4      &  $<$25.5              &$<$19.5           &   \\
36 & 14.2 & {\em XMM--N}  & 9.1$\pm$4.1  &  5.2$\pm1.3$          &5.2   $\pm$ 2.4   & \\
37 & 4.7  & \chandra~     & $<$14.2      &  $<$31.9              &$<$6.3            &   \\                                
38 & 17.6 & {\em XMM--N}  & $<$3.0       & $<$4.3                &$<$4.8            &           \\
39 & 21.0 & {\em XMM--N}  & 37.0$\pm$4.8 & 32.5$\pm9.7$          &41.7  $\pm$ 5.4   &\\
40 & 21.1 & {\em XMM--N}  & 30.6$\pm$4.3 & 20.0$\pm2.1$          &14.1  $\pm$ 2.0   & \\
41 & 26.8 & {\em XMM--N}  & 21.9$\pm$3.4 & 25.1$\pm5.4$          &45.7  $\pm$ 7.1   & \\
42 & 14.5 & {\em XMM--N}  & $<$2.0       & $<$2.7                &  $<2.5$          &    \\
43 & 14.1 & {\em XMM--N}  & 46.3$\pm$6.8 & 20.4$\pm3.0$          &38.0  $\pm$ 5.6   & \\
44 & 22.0 & {\em XMM--N}  & 49.7$\pm$5.4 & 26.5$\pm4.2$          &19.1  $\pm$ 2.1   & \\
\hline                                    
\end{tabular}\end{center}         
Column: 
(1) ID ; 
(2) Total effective exposure in ks;
(3) Telescope (\xmm~or \chandra~);
(4) Background$-$subtracted count rate in the T band in units of $10^{-4}$ Count/s; 
(5) Flux in the T band in units of 10$^{-15}$ erg s$^{-1}$ cm$^{-2}$;       
(6) Logarithm of the F$_{2-10}$/F$_{R}$ ratio.
\end{table*}

\subsection{MIR properties of  \xmogs}

We computed 5.8 $\mu$m monochromatic luminosities of our sources using a linear interpolation between the observed 8 $\mu$m  and 24 $\mu$m flux densities at the wavelength corresponding to 5.8 $\mu$m at the source rest-frame (e.g. Fiore et al. 2008a for further details).
The F$_{24\mu m}$/F$_R$ values as a function of the $\lambda L_{5.8 \mu m}$ luminosity for the SWIRE $F_{24\mu m} > $ 1 mJy sample and for our \xmog~ sample are compared in Fig. 3a. 

It can be seen that our selection criteria (F$_{24\mu m}$/F$_R>$ 2000 and F$_{24\mu m}>1.3$ mJy) are well suited to find highly obscured, high--luminosity 
($\lambda L_{5.8 \mu m}$ $\geq 10^{44}$ \ergs) sources.  Furthermore, the combination of large F$_{24\mu m}$ and extreme MIR colors guarantees that the vast majority of our EDOGs are AGN--dominated in the MIR band (e.g. Polletta et al. 2008a, Dey et al. 2008, Donley et al. 2008). 
Our finding is therefore in agreement with the correlation between F$_{24\mu m}$/F$_R$ and MIR luminosity reported by Fiore et al. (2008a) for obscured AGNs (see their Figs. 2a and 2b).
  
On the other hand, highly luminous (i.e. $\lambda L_{5.8 \mu m}$ $> 10^{44}$ \ergs) MIR sources 
with a F$_{24\mu m}$/F$_R$ value in the range 10-100  predominantly show a  continuum-dominated MIR spectrum with PAH 
emission features, suggesting they are composite ULIRG systems containing both a 
dominant AGN and a starburst component  (Sajina et al. 2007, Yan et al. 2007). MIR-selected, 
luminous Type 1 quasars typically fall in this region of the F$_{24\mu m}$/F$_R$ - $\lambda L_{5.8 \mu m}$ diagram (e.g. Brand et al. 2007; Fiore et al. 2008b; Polletta et al. 2008a).
Finally, sources at lower luminosities (i.e. $\lambda$ $L_{5.8 \mu m} <$ 10$^{44}$ \ergs) can be likely associated with Seyfert-like AGNs (obscured and unobscured) and starburst galaxies.

In Fig. 3b is shown the distribution of the $\lambda L_{5.8 \mu m}$ luminosities as a function of the redshift for our \xmogs~ sample.
 For comparison we also plot the sources from the CDFS and COSMOS fields (Fiore et al. 2008a, 2008b) with F$_{24\mu m}$/F$_R>$2000.
The combination of our selection criteria with the large area of the SWIRE survey allows us to collect the most IR luminous sources at all redshifts: objects in our \xmog~sample are, on average, \simgt0.5 (\simgt1.5--2) order of magnitude brighter than those selected in the  2 ($<$0.1) deg$^{2}$ COSMOS (CDFS) survey.
Assuming that \xmogs~host an AGN at their center, it is possible to give a rough estimate of the  bolometric luminosity ($L_{Bol}$) using the relationship {\it log}$L_{Bol}$ = {\it log}$\lambda L_{5.8 \mu m}$ $+$ 1.16 from Elvis et al. (1994). In particular, Polletta et al. (2008a) point out that this relationship likely provides only a lower limit on real value of $L_{Bol}$. By proceeding in this way we obtained values of  bolometric luminosities  in the range  10$^{12}$\simlt~$L_{Bol}$\simlt~10$^{14}$ $L_{\odot}$. \xmogs~are therefore placed at the  brightest end of the bolometric luminosity function of galaxies in the Universe.

\section{X-ray observations}
\label{sec:x-ray}

\subsection{Data reduction}

We searched for all \chandra~and \xmm~archival observations available for the SWIRE fields listed in Sect. 2.1 as of February 2008.
The X-ray coverage of the sources in our sample is highly inhomogeneous, ranging from shallow 5 ks \chandra~observations of the $Bo\ddot{o}tes/NOAO$ field, up to medium-deep observations in the \xmm~SXDS field ($\sim$50 ks) and in the \chandra~Lockman Hole region ($\sim$70 ks).
Twenty-one sources fall within the FOV of an \xmm~observation, while the sources with \chandra~data are twenty-three.

Event files of the \chandra~observations were retrieved from the \chandra~X-ray Center (CXC) via the Web ChaSeR (Chandra Archive Search and Retrieval) interface.
The data reduction was performed following the standard procedures outlined in the Science Analysis Threads for ACIS data at the CIAO Web site\footnote{\textit{http://cxc.harvard.edu/ciao}}.

Standard \xmm~SAS tasks \textit{epproc} and \textit{emproc} (SAS v. 7.1.0)\footnote{\textit{http://xmm.esac.esa.int/sas/}} were used to linearize the \pn~and \mos~event files.
The event files were processed using  the latest calibration files
and cleaned up from effects of hot pixels and  cosmic rays contamination.
X-ray events corresponding to patterns 0--12 (0--4) for  the \mos (\pn) cameras were selected.
\xmm~observations sometimes experience high background flaring periods due to the crossing of clouds of soft (i.e. E$<$1 MeV) protons, produced by the sun and projected towards the Earth in the solar wind, that are accelerated by magnetic reconnections in the magnetosphere.
In order to produce a cleaned event file we removed all the high background periods identified in
light curves at energies $>$10 keV, where the contribution from the emission from real X-ray sources is minimized.

\begin{figure*}
\begin{center}
\includegraphics[width=8cm,height=8cm]{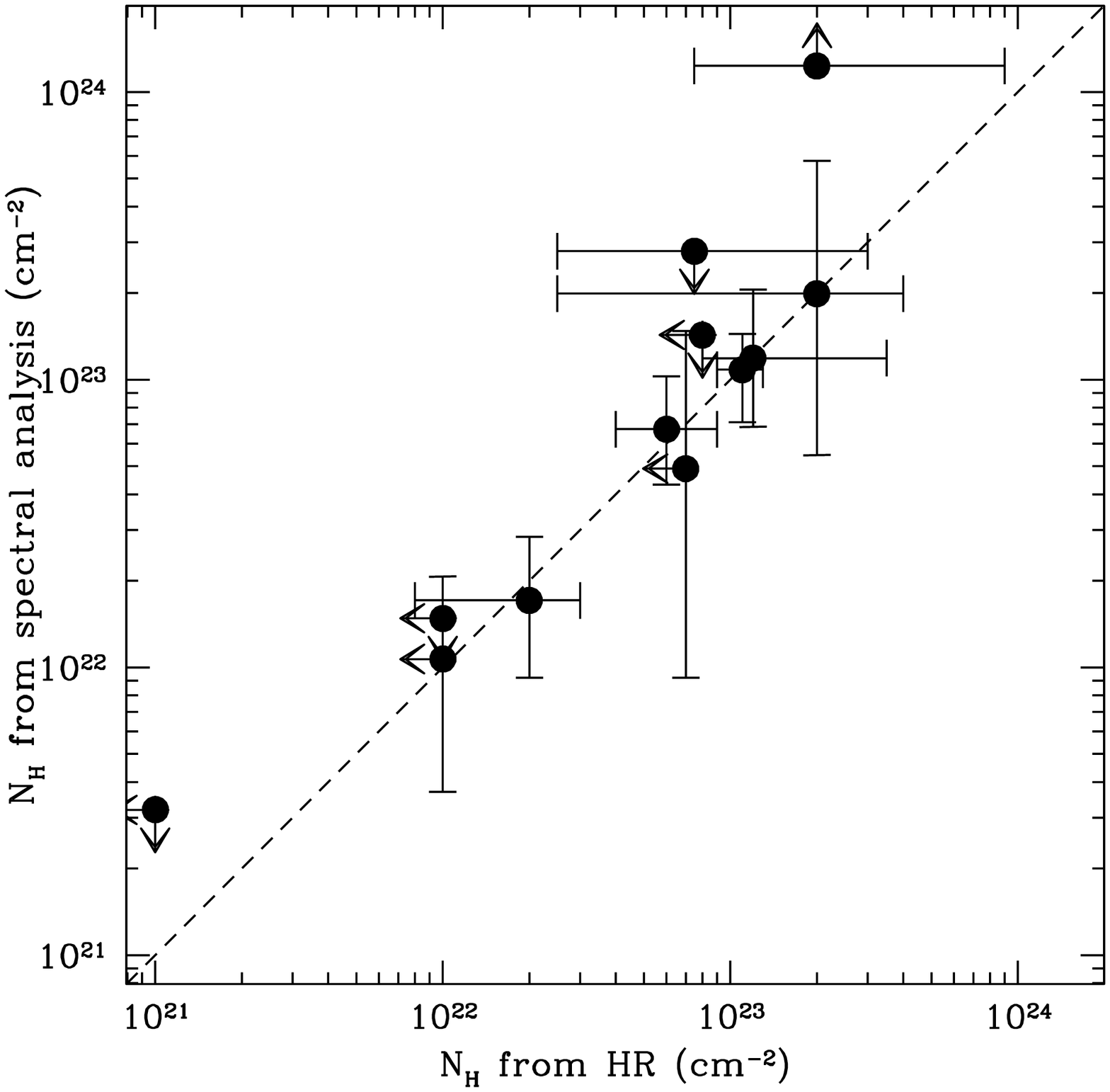}\hspace{1cm}\includegraphics[width=8cm,height=8cm]{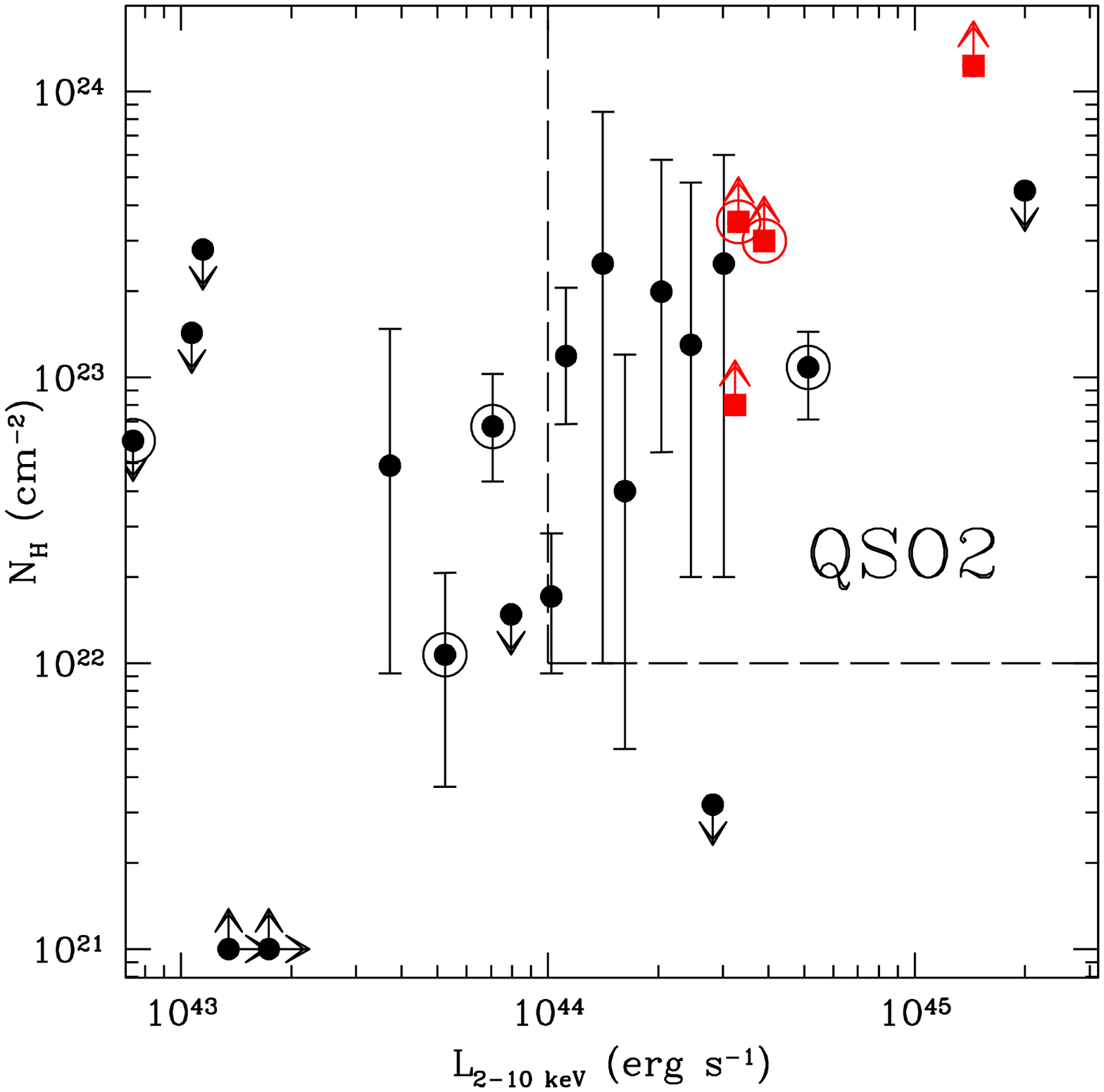}
\caption{{\it (a) Left panel:} Comparison of the best fit values of \nh~to those obtained with HR technique, for the 12 sources with available spectral analysis. {\it (b) Right panel:}  Distribution of column density \nh~as a function of the \lum. Filled squares indicate sources consistent with being Compton Thick AGNs. For these sources we plot the lowest value at 90\% confidence level for the \nh~parameter. Sources with a detection only in the Hard band have a lower limit both in the \nh~and \lum~values. Circled points represent the spectroscopically-identified sources. Dashed line marks the locus for sources with \nh~and \lum~values typical of  QSO2s (\nh$>10^{22}$ cm $^{-2}$, \lum$>10^{44}$ erg s$^{-1}$).}
\end{center}
\end{figure*}

\subsection{X-ray Fluxes}
\label{xray}
The count-rate in the 0.5-10 keV band (indicated as T band hereafter) was obtained by the command  \textit{sosta} in the  Ximage\footnote{ \textit{http://heasarc.gsfc.nasa.gov/xanadu/ximage/ximage.html}} package.
\begin{table*}
\caption{X-ray spectral properties of the 23 sources showing a detection in at least one of the two energy bands, S and H.}
\label{tab:xray}
\begin{center}
\begin{tabular}{cccccccccc}
\hline\hline\\
\multicolumn{1}{c} {ID}&
\multicolumn{1}{c} {CR$_{0.5-2}$}&
\multicolumn{1}{c} {CR$_{2-10}$}&
\multicolumn{1}{c} {HR}&
\multicolumn{1}{c} {Flux$_{2-10}$}&
\multicolumn{1}{c} {N$_H$}&
\multicolumn{1}{c} {log(L$_{X,obs}$)}&
\multicolumn{1}{c} {log(L$_{X}$)}&            
\multicolumn{1}{c} {log($\lambda L_{5.8}$/\lum)}&
\multicolumn{1}{c} {Analysis} \\
(1) & (2) & (3) & (4) & (5) & (6) & (7) & (8) & (9) & (10) 
 \\\hline\\ 
1 & 19.4 $\pm$3.4 & 6.1 $\pm$2.5  &$-$0.52$\pm$0.19&10.2$_{-4.0}^{+3.8}$   &1.1$_{-0.7}^{+1.4}$      & 43.68 & 43.72        &     1.4 1 &SP\\
4 & 1.7  $\pm$0.8 & 2.2 $\pm$0.9  & 0.13$\pm$0.31  &8.5$_{-4.2}^{+5.9}$    &19.9$_{-14.4}^{+37.8}$   & 43.90 & 44.31        &     1.31  &SP\\ 
5 & 2.7  $\pm$1.3 & 1.8 $\pm$1.2  &$-$0.25$\pm$0.53&2.2$_{-0.6}^{+2.8}$    &$<$14.3                  & 42.93 & 43.03        &     2.1   &SP\\ 
6 & 21.1 $\pm$2.5 & 63.7$\pm$4.3  & 0.49$\pm$0.07  &113.1$_{-30.0}^{+17.0}$&10.8$_{-3.8}^{+3.6}$     & 44.49 & 44.71        &     0.95  &SP\\ 
7 & 4.0  $\pm$2.8 & 7.4 $\pm$3.9  & 0.30$\pm$0.44  &14.2$_{-3.0}^{+3.0}$   &25.0$_{-23.0}^{+35.0}$   & 44.40 & 44.48        &     1.43  &HR\\ 
8 & $<$6.9        & 4.8  $\pm$4.1 & $>$-0.53       &3.4$_{-1.6}^{+1.6}$    &$>$0.1                   & 43.13 & $>$43.13$^a$ &     1.29  &HR\\ 
9 & 1.8  $\pm$1.6 & 2.8 $\pm$1.9  & 0.21$\pm$0.54  &5.3$_{-1.9}^{+1.9}$    &25.0$_{-24.0}^{+60.0}$   & 44.08 & 44.15        &     1.47  &HR\\ 
10& 1.0  $\pm$0.9 & $<$1.5        & $<$0.25        &$<$2.3$_{-1.4}^{+1.4}$ &$<$6.0                   & 42.82 & 42.87        &     1.76  &HR\\ 
11& 5.0  $\pm$2.3 & 3.5 $\pm$2.4  &$-$0.18$\pm$0.39&5.8$_{-2.7}^{+2.6}$    &4.9$_{-4.0}^{+9.9}$      & 43.40 & 43.57        &     1.73  &SP\\ 
12& 1.9  $\pm$1.6 & 5.7 $\pm$2.5  & 0.50$\pm$0.44  &27.1$_{-3.9}^{+4.8}$   &267.9$_{-144.8}^{+274.7}$& 43.40 & 45.16        &     0.34  &SP\\ 
19& 2.2  $\pm$1.3 & 8.6 $\pm$4.4  & 0.60$\pm$0.62  &12.6$_{-7.9}^{+7.9}$   &25.0$_{-17.0}^{+75.0}$   & 44.10 & 44.92        &     1.76  &HR\\ 
21& 13.0 $\pm$5.3 & 10.6 $\pm$4.9 &$-$0.10$\pm$0.31&24.8$_{-6.4}^{+6.4}$   &4.0$_{-3.5}^{+8.0}$      & 44.19 & 44.21        &     1.05  &HR\\ 
22& 5.9  $\pm$3.0 & 6.7 $\pm$3.7  & 0.06$\pm$0.37  &11.3$_{-4.2}^{+4.2}$   &13.0$_{-11.0}^{+35.0}$   & 44.35 & 44.39        &     1.56  &HR\\ 
25& 10.9 $\pm$3.7 & 4.5  $\pm$5.0 &$-$0.42$\pm$0.43&18.6$_{-3.8}^{+3.8}$   &$<$45.0                  & 45.26 & 45.30        &     1.33  &HR\\ 
31& $<$2.2        & 0.9  $\pm$0.4 & $>$-0.15       &1.7$_{-0.7}^{+0.7}$    &$>$0.1                   & 43.24 & $>$43.24$^a$ &     1.70  &HR\\ 
32& 0.7  $\pm$0.3 & 1.8 $\pm$0.5  & 0.49$\pm$0.31  &3.1$_{-0.5}^{+0.5}$    &75.0$_{-40.0}^{+75.0}$   & 44.10 & 44.53        &     2.23  &HR\\ 
33& 0.7  $\pm$0.3 & 1.9 $\pm$0.6  & 0.48$\pm$0.28  &3.8$_{-0.6}^{+0.6}$    &60.0$_{-30.0}^{+45.0}$   & 44.10 & 44.59        &     1.82  &HR\\ 
36& 3.7  $\pm$2.5 & 9.4 $\pm$3.8  &  0.43$\pm$0.40 &4.3$_{-1.2}^{+6.6}$    &$<$28.0                  & 43.01 & 43.06        &     1.21  &SP\\ 
39& 11.2 $\pm$2.8 & 25.9 $\pm$3.9 & 0.40$\pm$0.14  &29.9$_{-13.4}^{+7.4}$  &6.7$_{-2.4}^{+3.5}$      & 43.65 & 43.85        &     1.03  &SP\\ 
40& 23.2 $\pm$3.5 & 6.9  $\pm$2.3 &$-$0.53$\pm$0.16&9.3$_{-1.3}^{+2.1}$    &$<$1.5                   & 43.88 & 43.90        &     1.34  &SP\\ 
41& 4.6  $\pm$1.9 & 17.5$\pm$2.8  & 0.58$\pm$0.18  &23.7$_{-8.1}^{+6.9}$   &11.9$_{-5.0}^{+8.7}$     & 43.88 & 44.05        &     0.46  &SP\\ 
43& 40.7 $\pm$5.8 & 3.3  $\pm$3.0 &$-$0.81$\pm$0.18&12.0$_{-2.7}^{+6.6}$   &$<$0.3                   & 44.44 & 44.45        &     1.04  &SP\\ 
44& 31.2 $\pm$4.1 & 18.4 $\pm$3.6 &$-$0.26$\pm$0.11&19.5$_{-7.5}^{+6.9}$   &1.7$_{-0.8}^{+1.1}$      & 43.94 & 44.01        &     1.24  &SP\\ 
\hline                                    
\end{tabular}\end{center}            
Column: 
(1) ID number; 
(2) Count rate in the S band in units of $10^{-4}$ count/s;
(3) Count rate in the H band in units of $10^{-4}$ count/s;
(4) Hardness Ratio between the H band and the S band computed with the formula H-S/H+S; 
(5) H band flux in units of $10^{-15}$ erg s$^{-1}$ cm$^{-2}$;
(6) Rest-frame column density \nh~in units of $10^{22}$ cm$^{-2}$;
(7) Logarithm of the observed luminosity in the H band (erg s$^{-1}$);                                 
(8) Logarithm of the absorption-corrected luminosity in the H band (erg s$^{-1}$);
(9) Logarithm of the $\lambda L_{5.8 \mu m}$/\lum~ratio;
(10) Analysis method: SP for spectral fitting, HR for hardness ratio. For \chandra~data the 2$-$10 keV count rate and flux was  extrapolated from the 2$-$8 keV value assuming $\Gamma$=1.9. \\
$^a$ For these sources we computed the absorption-corrected luminosity using the lower limit on the \nh~value. Accordingly, the intrinsic luminosity shown here is also a lower limit. 
\end{table*}

Using this command, we derived the number of source counts within a box region with a 5 (15) arcsec side
for \chandra~(\xmm) data centered on the position of the optical counterpart of the X-ray source and 
 also corrected them for the effects due to the mirror vignetting and the off-axis point spread function degradation.
We  repeated the same procedure to estimate the background by the extraction of the background counts from a source-free region around the target. We were thus able to infer the background-subtracted source count rate and its statistical significance. A 3$\sigma$ upper limit is automatically calculated by {\it sosta} if the detection significance is less than 10$^{-3}$. According to this procedure, we obtained 25 positive detections and 19 upper limits in the T band for the 44 sources in our sample.
     
The count rate was subsequently converted into flux using the PIMMS\footnote{\textit{http://heasarc.nasa.gov/Tools/w3pimms.html}} tool. As underlying spectral model we assumed a power law (with a photon index fixed to $\Gamma$ = 1.9) modified by absorption from neutral gas with a rest-frame column density of \nh=10$^{23}$ cm$^{-2}$. This is an intermediate \nh~value for F$_{24\mu m}$/F$_R$ selected DOGs (see below).
     
Table 2 lists the resulting T band count rate and flux along with the effective exposure time for each source in the sample. The latter has been obtained from the nominal exposure time of the observation after correction for the vignetting factor at the source position on the detector.
We found that $\sim$45\% of the sources have an effective exposure time that is $\geq$15 ks. 
Fluxes in the T band span from 1.8$\times10^{-15}$ to 1.1 $\times10^{-13}$ erg s$^{-1}$ cm$^{-2}$
with a distribution mean value of $\sim$ 2 $\times10^{-14}$ \cgs.

\section{X-ray spectral properties}
\label{spectral}
     
In order to characterize their spectral properties (with particular emphasis on the absorbing column density \nh), we selected the sources showing a detection in at least one of the following energy bands: soft (i.e. 0.5-2 keV, S band hereafter) or hard (i.e. 2-10 keV, H band). 
This criterion indeed allows to derive the hardness ratio (HR)  and, therefore, provide an estimate 
of the absorption column density even if just a handful of counts are collected during an observation.
This condition was satisfied by 23 out of 44 sources, i.e. $\sim$50\% of the sample and,
for the 12 brightest sources, i.e. with $\geq$40  net counts in the T band, we were also able to perform a basic spectral analysis (we refer hereafter to this sample of 23 \xmogs~ as the 'X-ray sample').

\subsection{Spectral Analysis}
 \label{spectralanalysis}    

All sources detected with a number of T band counts $\geq$40 are observed with \xmm. 
Spectra were acquired from the cleaned event files of each \epic~camera, i.e. \pn, \mosuno~and \mosdue. 

Source counts were extracted from circular regions with a radius in the range of 10--20 arcsec. The radius was chosen in order to optimize the signal-to-noise ratio (S/N) for each source, but in same cases it was limited by the presence of nearby sources or 
CCD gaps. 
Appropriate response and ancillary files for all the \epic~cameras were created using RMFGEN and ARFGEN tasks in the \xmm~SAS, respectively.
In case of multiple observations of the same source, we combined the source and the background spectra from different exposures using the tool \textit{mathpha} in order to improve the photon statistics. 
In doing so we also computed the averages of Response Matrix Files and Ancillary Region Files, weighted for exposure times, using the tools \textit{addrmf} and \textit{addarf}.
We performed the spectral analysis for 12 sources having net T band counts in the range 40--600 (with  a median value of $\sim$80 counts). All the fits discussed in this paper include the absorption due to the line-of-sight Galactic column density (Dickey \& Lockman 1990).

For the nine sources with a number of counts $<$150 we used the C-statistic minimization method (Cash 1979),
being particularly well-suited for  low--count spectra.
 This is a maximum likelihood method that allows the use of unbinned data. Since it assumes that the counts collected in a given channel follow a Poisson distribution, it cannot be applied to background-subtracted data. Therefore, before performing the spectral modeling we accurately parametrize the background spectrum, extracted from a large source-free region for each source, with a model consisting of a broken power-law plus Gaussian lines, as suggested by Lumb et al. (2002). Then we included this background model in the source spectral fit (fixing the values of the parameters and rescaling the normalizations to the area of the source extraction region). 
We assumed an absorbed  power-law  as spectral fitting model.
Due to the low number of counts we decided to fix the power-law photon index $\Gamma$ to 1.9, i.e.  the average value found for large samples of quasars in the 2-10 keV band (e.g. Reeves \& Turner 2000; Piconcelli et al. 2005 and references therein)

For the remaining three sources (namely source Nos. 6, 39, 44) with $>$ 150 net spectral counts  we were able to perform the spectral analysis using the $\chi^2$ minimization method. 
In particular, the large number of collected counts for the source No. 6 (SWIRE2 J021749.00$-$052306.9) allowed to significantly detect an Fe K$\alpha$ emission line at $\sim$6.4 keV in this high-luminosity (\lum$\approx$ 5 $\times$ 10$^{44}$ \ergs) obscured quasar at $z \sim$ 1 (see Sect. 6.2  for further details).
Table 3 reports \xr~ counts and spectral parameters for these 12 sources (labeled with SP). We also report the observed \lum, the absorption corrected \lum~and the ratio $\lambda L_{5.8 \mu m}$/\lum~value for each of them.

\subsection{Hardness Ratio Analysis}

For the 11 sources detected with $<$40 T band counts  we calculated the background-subtracted count-rate in the  S and H band. 
We computed the HR value, defined by the formula C$_{H}$-C$_{S}$/C$_{H}$+C$_{S}$, where C$_{S}$ (C$_{H}$) is the net count-rate in S (H) band.
The HR provides an indication about the flatness of an X-ray spectrum and, therefore, it can be used for an estimate of the strength  of X-ray obscuration, since the photoelectric absorption mainly affects the soft X-ray portion of the spectrum (i.e. photoelectric absorption cross-section $\sigma (E)\propto E^{-3.5}$).

To evaluate the absorbing column density \nh~, we created a simulated spectrum adopting as model  a power law (with photon index $\Gamma$ = 1.9) modified by a rest-frame absorption. We varied the column density \nh~in the range 20$<$log(\nh)$<$24 \cm2~by steps of $\Delta$log(\nh)=0.2, and the redshift in the range 0$\leq$z$\leq$3 by steps of $\Delta z$=0.5. By computing the HR of the simulated spectrum for each value of \nh~and z, we created a set of curves corresponding to different values of \nh~in the plane $z$ versus HR that allowed us to evaluate the absorption \nh~for each source.
Thus, we were able to compute both the observed and the absorption corrected \lum. 
For 2 source (Nos. 8 and 31) with a positive detection only in the H band, we obtained a lower limit on HR and, hence, \nh. The latter was used in the computation of the absorption corrected luminosity
Table 3 reports \xr~counts and spectral parameters for these 11 source (labeled HR).

Previous works (e.g. Perola et al. 2004) suggested that the \nh~values obtained through the HR technique tend to be overestimated.
In order to check the reliability of the HR technique in case of a small number of counts, we compared the \nh~inferred by spectral analysis with the \nh~value obtained using the method outlined above for the subsample of 12 sources (see Sect. \ref{spectralanalysis}). The  values of \nh~determined for each sources are in good agreement within errors as shown in Fig. 4a. Notably, the only source  whose \nh~estimate does not match the one-to-one relationship is the Compton-thick QSO2 candidate SWIRE2 J022003.95-045220.4 (i.e. source No. 12, see Appendix for further details).
This suggests that, due to the intrinsically different spectral shape of Compton-thick sources, the HR analysis can lead to underestimate the \nh~value in these sources.

\begin{figure*}
\begin{center}
 \includegraphics[width=8cm,height=8cm]{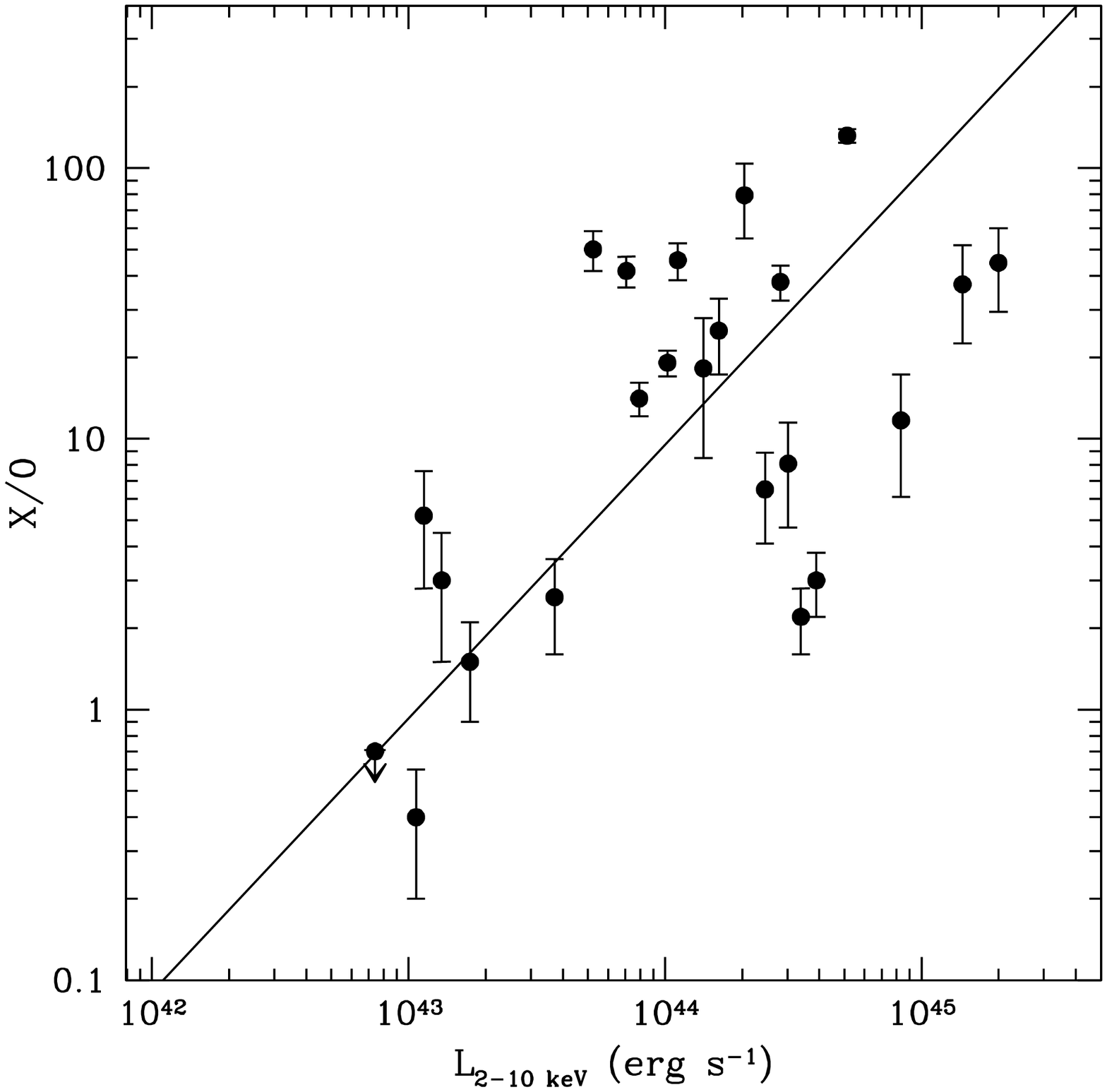}\hspace{1cm}\includegraphics[width=8cm,height=8cm]{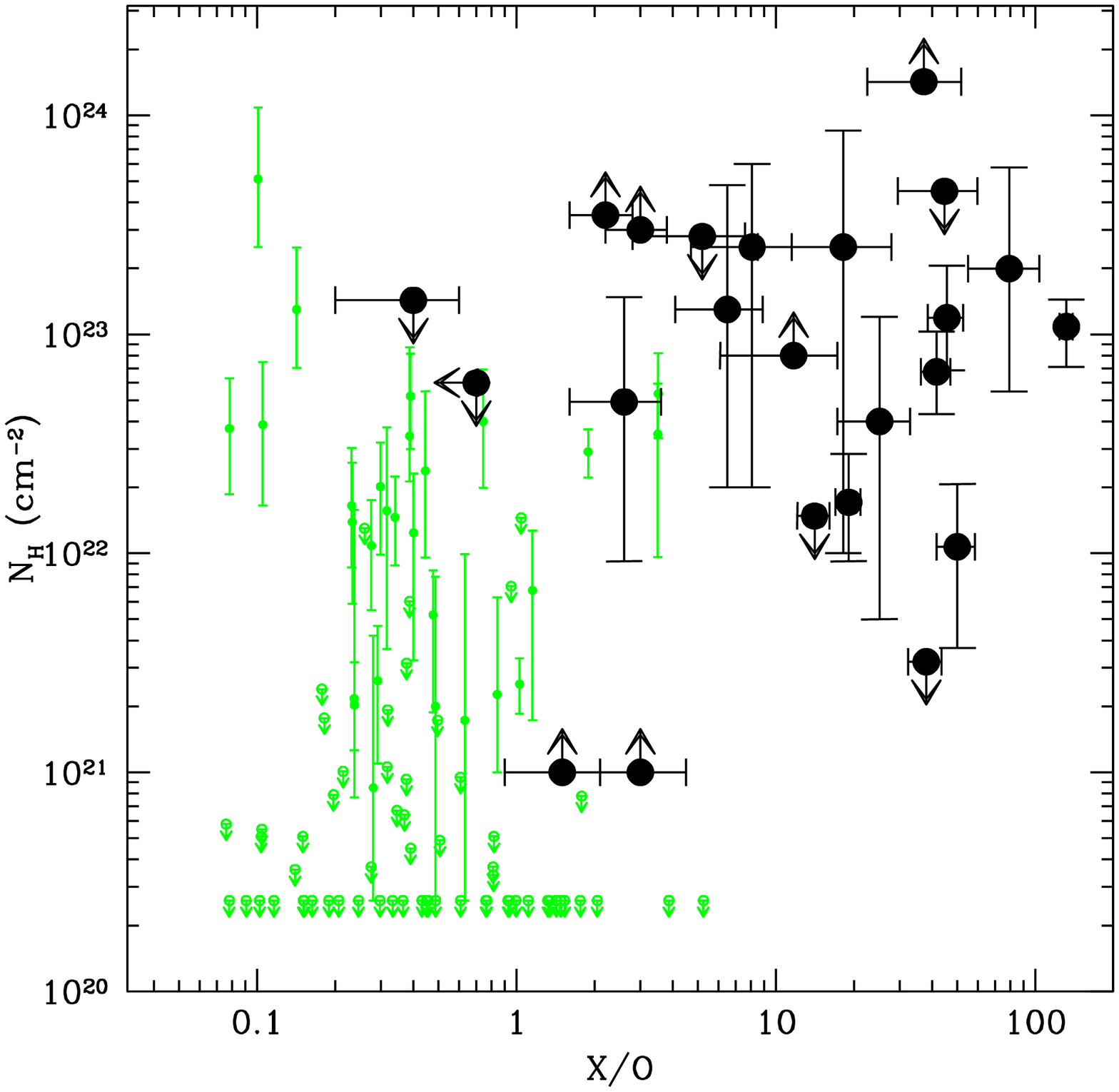}
 \caption{{\it a) Left panel:} F$_{2-10}$/F$_R$ ratio as a function of the 2-10 keV luminosity for the \xr~sample. The solid line is the best log(F$_{2-10}$/F$_R$)-log(L$_{2-10 kev}$) linear regression obtained by Fiore et al. (2003) for a  large sample of optically obscured AGNs (see Sect. 4 for details). {\it b) Right panel:} F$_{2-10}$/F$_R$ ratio versus column density \nh~for the sources in the \xr~sample (large points). For comparison we also plot the values obtained for a sample of the 99 brightest, hard X-ray selected sources in the wide-area (6 deg$^2$) XMM-LSS survey from Garcet et al. (2007) (small circles).}
\label{xo}
\end{center}
\end{figure*}

\subsection{Absorption properties}

All the sources in the \xr~sample show a hard \xr~luminosity value typical of AGNs, i.e. \lum$\gg10^{42}$ \ergs.
We found that the vast majority (i.e.$\sim$95\%) of the 23 \xmogs~ with X-ray spectral data are consistent with being obscured by neutral gas with an intrinsic column density of \nh$\geq$ \e22~\cm2~(Fig. 4b). Unfortunately, the combination of large $z$ values and, mostly, the limited photon statistics do not allow us to test the presence of warmly ionized absorbing gas in our sources.
In particular the fraction of X-ray obscured \xmogs~showing a cold absorption column density strictly larger than \e22~\cm2~is $\sim$55\%. The uncertainties on this fraction is due to the large errors on the \nh~value for the sources detected with a small number of counts.
Table~\ref{tab:summary} lists the source breakdown according to the level of absorption. This table, as well as Figure 4b, also reveals the common presence ($\sim$55\%) of QSO2s among the \xmogs~in the X-ray sample. This finding highlights the efficiency of our MIR selection criteria in collecting a large number of these elusive AGNs, which are easily missed in optical and X-ray surveys. This makes promising their application to large samples of X-ray well-exposed \xmogs~with the goal of studying the properties of the absorption  at the brightest end of the AGN X-ray luminosity distribution.
It is also worth noting that the sources with the highest absorption column densities fall within the QSO2 locus in  the \nh~versus 2-10 keV luminosity plane.

To further test the real capability of our MIR selection criteria in gathering QSO2s, we compared these results with those obtained for
 a control sample consisting of 20 randomly-selected sources from the same SWIRE fields taken into account in our work. 
We picked out these sources for having a hard X-ray flux range (\fhx$\approx$10$^{-15}$--10$^{-13}$ erg cm$^{-2}$ s$^{-1}$) and a redshift distribution similar to those of our X-ray sample, regardless of their MIR and optical properties.
Photometric redshifts for these sources are taken from Rowan--Robinson et al. (2008) (13 sources) or have been derived from optical and IR photometric data (7 sources). 
Spectral parameters were derived by the individual X--ray spectral analysis of each source.
The fraction of sources in the control sample with \nh~$\geq10^{22}$ cm$^{-2}$ is about 15\%. Finally, the fraction of QSO2s is $\sim$5\%.

\begin{table}
\caption{The source breakdown according to the absorption properties. See Sect. 4.3 for details.}
\label{tab:summary}
\begin{center}
\begin{tabular}{lcc}
\hline\hline\\
\multicolumn{1}{l} {Source}&
\multicolumn{1}{c} {X-ray}&
\multicolumn{1}{c} {Control} \\
Type& Sample &  Sample \\ 
& (1) & (2)  \\
 \\\hline\\ 
\nh$<10^{22}$     &  13-52\%  & 85\% \\
\nh$\geq10^{22}$  & 48-87\%  & 15\% \\
QSO2s             & 43-60\%   &  5\% \\
N. of Sources      &  23       &  20 \\
\hline                                    
\end{tabular}\end{center}            
Column: 
(1) Percentage of objects for each source type found in our sample (i.e. for the absorbed sources we report the maximum  and minimum value derived assuming the best fit value and the lowest value at 90\% confidence level for the \nh~parameter, respectively);
(2) Percentage of objects for each source type in the control sample (20 objects).
\end{table}

\subsection{X-ray to Optical flux ratio}
 Optically-selected AGNs typically show an F$_{2-10}$/F$_R$ flux ratio, in the range 0.1--10, while $\sim$ 20\% of AGNs selected in hard \xr~deep surveys has an F$_{2-10}$/F$_R\geq$ 10 (see Alexander et al. 2001; Fiore et al. 2003). Follow-up studies of
these EXOs found that most of them are dust-enshrouded AGN at $z$\simgt~1 (e.g. Maiolino et al. 2006 and references therein). 
Fiore et al. (2003) discovered the existence of a striking correlation between  F$_{2-10}$/F$_R$  and 2-10 keV luminosity for non Type 1 AGNs whereby
the most luminous \xr~sources tend to show the highest F$_{2-10}$/F$_R$ values. 
This is due to the fact that Compton-thin  AGNs  have their nuclear optical/UV light strongly attenuated differently from the X-ray light and, therefore, their F$_{2-10}$/F$_R$ represents the ratio between the X-ray flux from the AGN and the host galaxy star light, which spans a small luminosity range.
Furthermore, the F$_{2-10}$/F$_R$ value is a proxy for the absorption in AGN (i.e. the higher is F$_{2-10}$/F$_R$, the higher is \nh) and, therefore, can be used to select obscured AGNs.

In order to investigate the effects of our MIR/Optical selection on the distribution of F$_{2-10}$/F$_R$, we computed this ratio for the sources in our sample\footnote{For the sources which are undetected in the 2-10 keV band we derived an upper limit in the F$_{2-10}$/F$_R$ value extrapolating the H band flux using as underlying
 spectral model an absorbed (\nh=10$^{23}$ \cm2) power law with $\Gamma$=1.9, e.g. Sect. 3.2.} (see Table 2).
In Fig.~\ref{xo}a we plot the distribution of F$_{2-10}$/F$_R$ as a function of the 2-10 keV luminosity for the \xr~sample.
A clear  correlation between F$_{2-10}$/F$_R$ and hard X-ray luminosity is found, even if characterized by a quite large scatter, in particular for the most X-ray absorbed sources. The solid line in the plot represents the best log(F$_{2-10}$/F$_R$)-log(L$_{2-10 kev}$) linear regression obtained by Fiore et al. (2003) from a large sample of \xr~selected, optically-obscured AGNs.

The F$_{2-10}$/F$_R$ value as a function of the column density \nh~for our \xr~sample is displayed in Fig.~\ref{xo}b. Notably, all but two sources with F$_{2-10}$/F$_R>$10 show X-ray absorption column densities consistent with being \nh \simgt~10$^{22}$ \cm2. We also report in this diagram the values obtained for a sample of the brightest 99 hard \xr~sources detected in the XMM-LSS survey (e.g. Garcet et al. 2007). This survey covers $\sim$ 6 deg$^{2}$ and, therefore, is one of the surveys with the widest coverage area in X-rays. This sample of 99 optically identified sources (half of sources show a $z>0.5$) covers roughly the same range of 2-10 keV flux (4$\times 10^{-15}< F_{2-10}< 2\times 10^{-13}$ ergs cm$^{-2}$ s$^{-1}$) of our X-ray sample, and the \nh~values were obtained from individual spectral analysis. From this figure it can be seen that \xmogs~have higher values of F$_{2-10}$/F$_R$ and that also the \nh~values are significantly higher, as expected on the basis of Fiore et al. (2003) and Cocchia et al. (2007) results.

\section{Discussion}
\label{discussion}

\subsection{Efficiency of our MIR selection criteria to collect QSO2s}
Our study was aimed at testing the capability of selection criterion (F$_{24\mu m}$/F$_R\geq$2000 and  F(24$\mu m$) $\geq$ 1.3 mJy)  
to isolate luminous, \xr~absorbed quasars at $z$ \simgt~1. 
The discovery of a QSO2 in $\sim$55\% of \xmogs~in the X-ray sample
is the most interesting result of our work (e.g. Table 4).
It is worth stressing that all the
\xmogs~with column densities consistent with being Compton-thick
 fall in the QSO regime, i.e. with a \lum$>10^{44}$ erg s$^{-1}$. 

We have thus tested the efficiency of our MIR technique
to recover a large number of these elusive objects missed in traditional 
surveys. They remained poorly-sampled even by
blind-searches performed in the deepest X-ray surveys
(e.g. Fabian 2002; Tozzi et al. 2006). Our choice of sampling at
F(24$\mu m$) \simgt~1 mJy appears crucial to identify very luminous
AGN-powered objects (as confirmed by the \lum~$>$10$^{43}$
\ergs~measured for the sources in the X-ray sample) and minimize any
strong contamination by non-AGN emission (e.g. Weedman et al. 2006;
Dey et al. 2008, Sacchi et al. in prep.).  

Our findings also highlights the excellent opportunity
represented by this simple approach in enabling many important studies
concerning the  nature of this dust-enshrouded, bright AGN population at high
redshifts. In particular, the systematic application of this method to wide-field MIR surveys with appropriately deep and 
homogeneous X-ray coverage will provide
useful insights on largely unexplored issues such as: (i) the demography
and surface density of highly obscured quasars; (ii) the multiwavelength
SED of distant QSO2s; (iii) the \nh~distribution at the largest X-ray luminosities; (iv) the luminosity
function of Compton-thin and Compton-thick quasars at $z>$ 1--2,
i.e. near the peak of the AGN activity.  Our work may be, therefore, a pathfinder study
for the next-generation observatories, when large areas of the sky will be surveyed at faint hard X-ray 
fluxes.

Our study represents the first attempt to characterize in details the
X-ray spectral  properties of \xmogs. As expected on the basis of
previous works based on analogous (although less severe) MIR selection
criteria (Fiore et al. 2008a, 2008b; Georgantopoulos et al. 2008; Polletta et
al. 2008a) these extremely red sources are found to harbor X-ray Type 2
AGNs. However, this MIR bright galaxy population deserves further and
deep investigations in X-rays since it constitutes by far  one of the
most intriguing recently-discovered class of extragalactic objects.
Extraordinary bolometric luminosities, high dust content and redshift
distribution of these sources lead to believe that the \xmog~phenomenon
can be potentially associated with the early stages of quasars
evolution as envisaged by Silk \& Rees (1998).
 
 Interestingly, as stressed by Dey et al. (2008), the space density of the  F(24$\mu m$) $\geq$ 1 mJy and F$_{24\mu m}$/F$_R\geq$1000
DOGs is comparable to that of unobscured quasars derived from the MIR luminosity function at $z$=2 (Brown et al. 2006).
The brightest DOGs likely are AGN--powered sources deeply obscured both in the optical and soft X-rays and, therefore, cannot be revealed by traditional optical/soft-X surveys, which miss a sizable fraction of the accretion history of the Universe.

In this paper we focus on the most extreme DOGs (representing $\sim$10\% of the F(24$\mu m$)$\geq$1 mJy and F$_{24\mu m}$/F$_R\geq$1000 DOGs) 
as a first step in the exploration of the X-ray spectral properties of this class of MIR bright, dust enshrouded galaxies. 
We indeed found that the vast majority of the \xmogs~harbor an obscured, intrinsically luminous \xr~AGN. It would be very interesting to extend this study to the whole DOG population, in order to provide an accurate estimate of the AGN space density at high z.

\begin{figure*} 
\label{fig:l58_lx}
\begin{center}
\includegraphics[width=9cm,height=9cm]{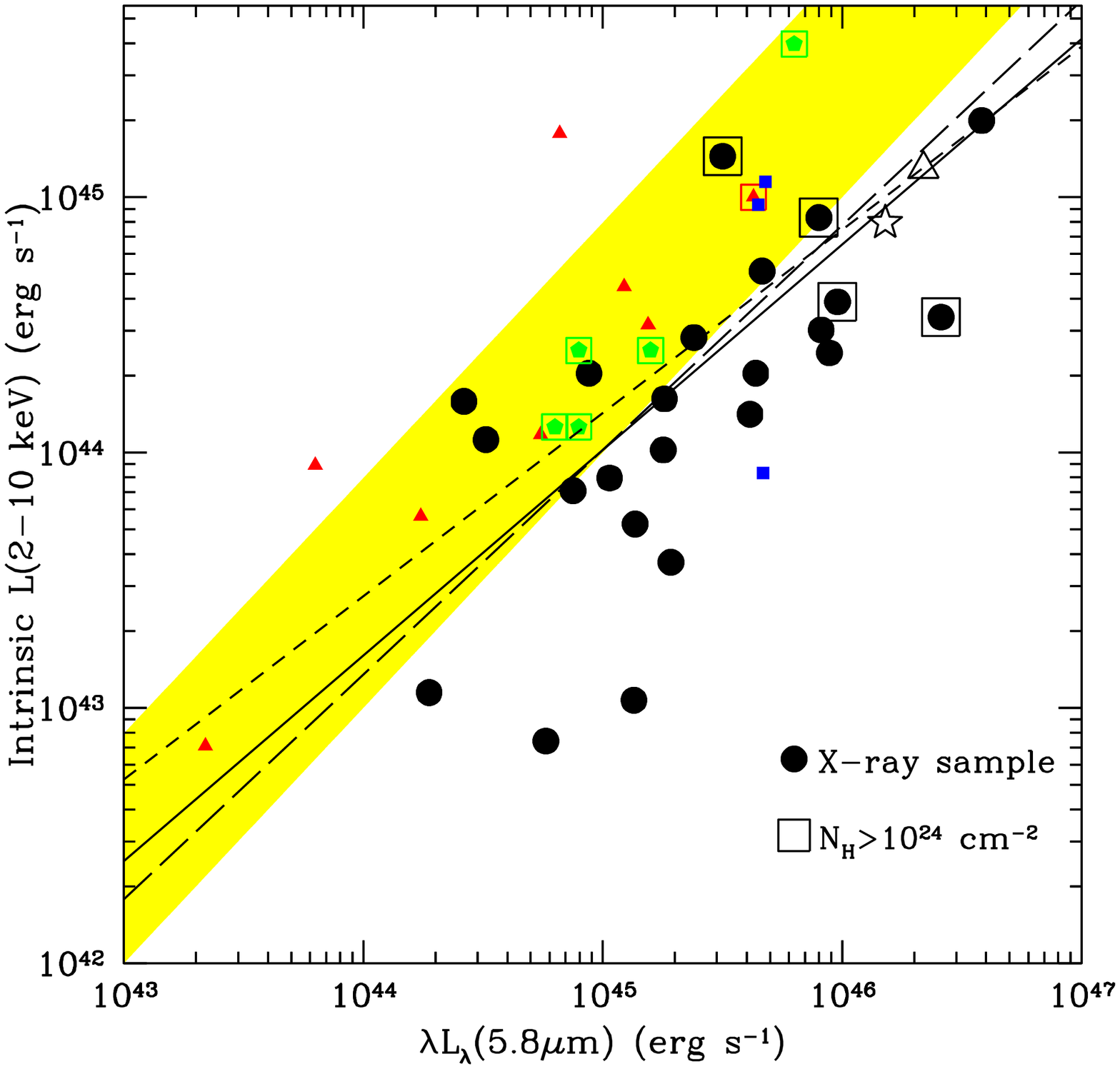}\hspace{0.3cm}\includegraphics[width=9cm,height=9cm]{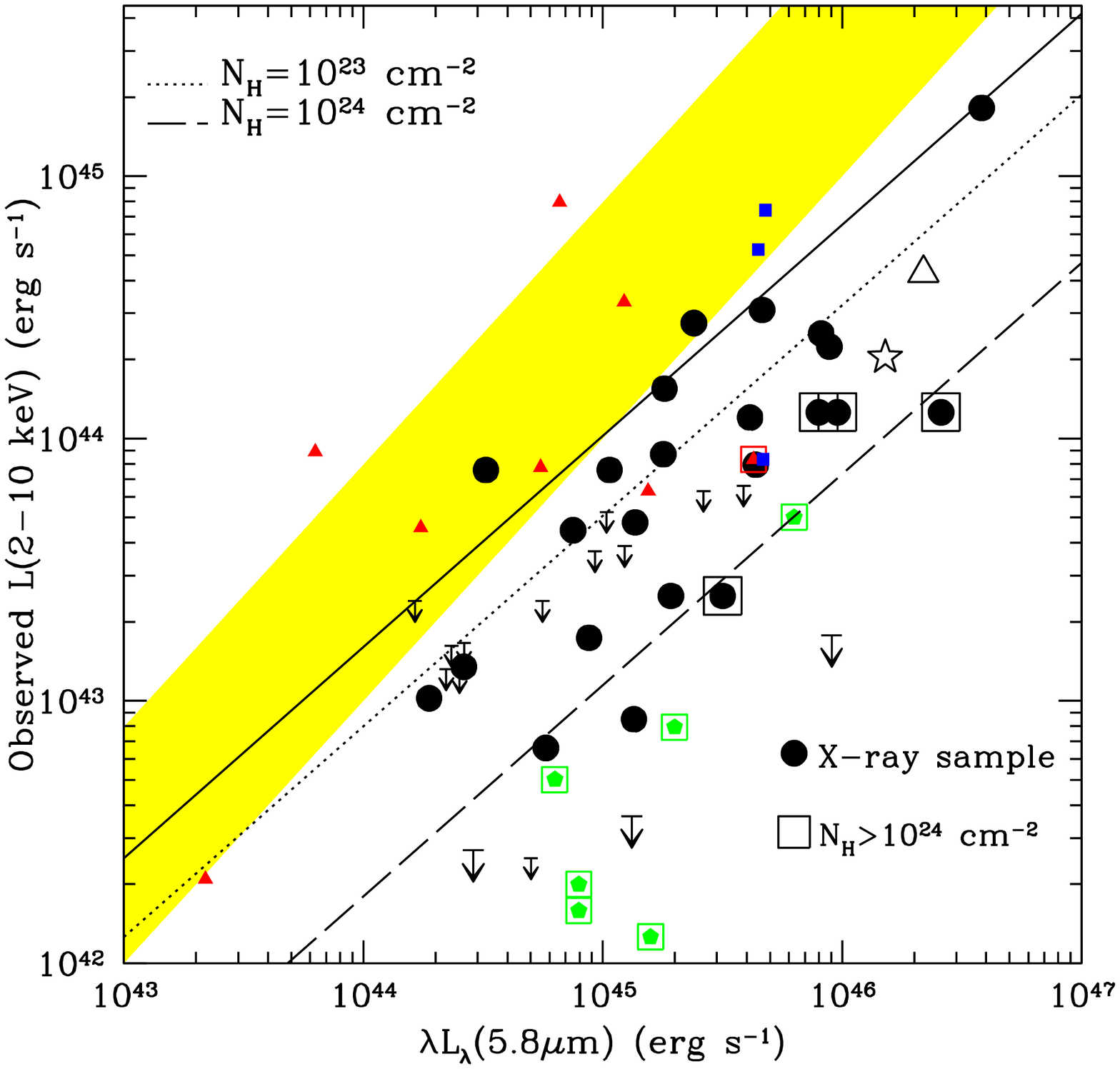}
\caption{{\it a) Left panel:} $\lambda L_{5.8 \mu m}$ versus \lum~for sources in our \xr~sample (filled circles), from Polletta et al. (2008a, filled squares), Alexander et al. (2008, filled stars), Sturm et al. (2006, filled triangles). Empty star represents the QSO2 IRAS 09104+4109 (Piconcelli et al. 2007a) and the empty triangle represents the QSO2 4C +24.36 (Johnson et al. 2007). See Sect. 5.2 for further details. Empty squares indicate sources absorbed by a column density of Log(\nh)$\geq$24 cm$^{-2}$. Solid line shows the linear regression obtained for sources with \lum~$\geq 10^{43}$ erg s$^{-1}$ in our \xr~sample. Short dashed line is the linear regression obtained from COSMOS Compton-thin AGNs (Fiore et al. 2008b). The long dashed line is the combination of the relationships
$\lambda L$(5100 $\AA$)/$\lambda L$(6.7$\mu$m) and \lum/$\lambda L$(5100 $\AA$) from Maiolino et al. (2007). 
{\it b) Right panel:} $\lambda L_{5.8 \mu m}$ versus the observed \lum. Downward-pointing arrows indicate the 3$\sigma$ upper limit on \lumo~for the \xr~undetected sources (large symbols indicate sources with exposures $\geq$15 ks, see Sect.~6.2). Other symbols are as in the previous panel. The solid line represents the linear regression obtained from our \xr~sample, while the dotted and dashed lines represent the ratio \lum$/\lambda L_{5.8 \mu m}$ expected for an absorption of \nh=$10^{23}$ and $10^{24}$ cm$^{-2}$, respectively. In both panels the shaded area indicates the range of \lum--$\lambda L_{5.8 \mu m}$ found for AGNs in the local Universe (Lutz et al. 2004).}
\end{center}    
\end{figure*}

\subsection{MIR luminosity versus Hard X-ray luminosity}

In Fig.~6a we compare the 2-10 keV luminosity with the 5.8 $\mu$m luminosity for sources  at the high--luminosity end of the AGN population.
Objects from our X-ray sample are indicated with filled circles, while the shaded area represents the region of \lum--$\lambda L_{5.8 \mu m}$ plain derived for AGNs in the local Universe by Lutz et al. (2004). In the figure are also plotted data taken from other samples of QSO2s available in the literature. In particular, we have included: (i)  three  QSO2s from the Polletta et al. (2008a) sample of  $z>$1 objects with F$_{24}>1$ mJy, F$_{24\mu m}$/F$_R>$400, $\lambda$ $L_{5.8 \mu m}>10^{12}$ L$_{\odot}$ and 10$^{21}$\simlt \nh \simlt10$^{23}$ \cm2~(namely LH\_02, LH\_A5 and LH\_A6; filled squares), (ii) six Compton-thick  $z \sim$ 2 QSO2 candidates from Alexander et al. (2008) which are extremely weak (or even undetected) in deep \xr~exposures but show evidence for strong AGN activity in the optical and/or MIR band (filled pentagons), and (iii) eight X-ray selected QSO2s ($L_{0.5-10}>$10$^{44}$ \ergs~and \nh~$>$10$^{22}$ \cm2) from the Sturm et al. (2006) sample (filled triangles).

We have also included in the plot two well-known, very powerful  ($L_{IR}$ $>$ 
10$^{12}$ $L_{\odot}$) dust-enshrouded QSO2s, i.e. IRAS 09104$+$4109 ({\it empty star}, $z$=0.442; e.g.
Hines \& Wills 1993; Piconcelli et al. 2007a) and 4C 23.56\footnote{We re-analyzed the \xmm~data of 4C 23.56. Our best-fit (an absorbed power law with 
\nh$=4.4^{+3.7}_{-2.4} \times 10^{23}$ cm$^{-2}$ and $\Gamma$ fixed to 1.9 plus a 
soft ($\Gamma\equiv$3) unabsorbed power law) is consistent with those reported 
by Johnson et al. (2006), confirming the QSO2 nature of this high-$z$ 
radiogalaxy. 
We derived a 2-10 keV intrinsic(observed) luminosity of \lum=1.3(0.4)$\times 10^{45}$ \ergs.} 
({\it empty triangle}, $z$=2.48; e.g. Johnson et al. 2006; Seymour et al. 2008).
Sources absorbed by a column density of Log(\nh)$\geq$24 \cm2~are marked with empty squares.

The solid line in this plot represents the linear regression derived for our sample:
\begin{equation}
Log(L_{2-10})=0.805 \times Log( \lambda\ L_{5.8\mu m}) + 7.785
\label{lbol_eq}
\end{equation}
once 3 sources with \lum$<10^{43}$ erg s$^{-1}$ were excluded, since their weakness in the hard X-ray band (see Table No. 3) suggests that their $\lambda L_{5.8 \mu m}$ is starburst--dominated.
 The long dashed line represents the result (i.e. Log(L$_{2-10}$)=0.88$\times$Log( $\lambda L_{6.7\mu m}$) + 4.42 ) obtained combining the relation between the rest-frame continuum luminosity 
$\lambda L_{5100 \AA}$ and the MIR luminosity $\lambda L_{6.7\mu m}$ with
the relation between \lum~and $\lambda L_{5100 \AA}$, both taken from Maiolino et al. (2007)
for a sample of 25 high luminosity (i.e. $\lambda  L_{5100 \AA} > 10^{46}$ \ergs) quasars at 2 $< z <$ 3.5.
The \lum--$\lambda L_{5.8 \mu m}$ relationship derived by Fiore et al. (2008b) 
 for a large sample of Compton-thin AGNs in C-COSMOS and CDFS surveys (i.e. $log(L_{2-10})=0.72\times log(\lambda L_{5.8 \mu m}$)+11.75) is marked with the short dashed line. 

As expected, we found a correlation between the two luminosities. In particular this relationship is not linear, and the $\lambda L_{5.8 \mu m}$/\lum~ratio increases with increasing $\lambda L_{5.8 \mu m}$. The shift at the highest MIR luminosities from the local Universe relationship suggests a possible evolution of the $\lambda L_{5.8 \mu m}$/\lum~with redshift.
Nonetheless it is evident from this plot that  there is a significant scatter in this relationship.
This probably means that we have put together a mix of different classes of sources, characterized by different properties and dominant energy mechanisms.
The line resulting from Eq.~\ref{lbol_eq} divides the $\lambda L_{5.8 \mu m}$--\lum~plane in two regions.
The region above this line includes all the X-ray selected QSO2s from Sturm et al. (2006) and the locus typical of AGNs in the local Universe,
which was derived by Lutz et al. (2004) from a sample of optically/X--ray bright Seyfert 2 galaxies. 
Selection effects therefore favor this region to be inhabited by the X-ray/optically 'loud' objects. All the sources from the Alexander et al. (2008) fall in this part of the diagram reflecting the fact that they are optically bright enough to be robustly identified via optical spectroscopy even if they likely are Compton-thick QSO2s. The intrinsic X-ray luminosities of these sources was indeed estimated on the basis of the luminosity of optical emission lines. Two out of three objects taken  from the Polletta et al. (2008a) sample showing moderate absorption in X-rays (\nh~\simlt~10$^{23}$ cm$^{-1}$) are located in this area. 
Interestingly, Sturm et al. (2006) emphasize the absence of dust emission and absorption features in the MIR spectra of their QSO2s unlike ULIRGs, which are very dusty systems with powerful circumnuclear starbursts. This points out that these highly luminous X-ray quasars were able to sweep away dust envelopes from their nuclear environment.

In the context of the  evolutionary sequence described in the previous Section, it seems possible that most of the objects falling in this region of the 
\lum--$\lambda L_{5.8 \mu m}$ plane have already passed through the initial dust-enshrouded phase of quasar activity/galaxy assembly (e.g. Silk \& Rees 1998) and are going to evolve into classical 'blue' QSOs traditionally collected in optical/UV surveys. IRAS 09104$+$4109 and 4C $+$24.36, which are respectively IR-- and radio--selected,  lies outside the locus of local Universe AGNs.

On the contrary, the majority of the sources located below the solid line in Fig.~6a might be objects with dust-absorbed powerful (ULIRG--like) nuclei in which the star-formation activity could significantly contribute to their MIR emission  or characterized by intrinsic extreme MIR/X luminosity ratios (e.g. Polletta et al. 2008b;  Piconcelli et al. 2009 in preparation). 
Accordingly, these sources may be the tip of a larger population representing the link between X-ray--selected QSO2s with featureless MIR spectra (e.g. Sturm et al. 2006) and the submillimeter galaxies (SMGs)  (e.g. Alexander et al. 2005, Pope et al. 2008b) which are well-known sites of strong star-formation activity.
In particular, these \xmogs~could belong to the minority fraction of SMGs  with optical-IR emission dominated by a dust-obscured quasar (e.g. Vignali et al. 2009) with MIR colors overlapping those of MIR power-law continuum AGNs  (Yun et al. 2008; Pope et al. 2008a). Thus, even if the position of a source in Fig.~6a strongly depends on the criterion of selection adopted, \xmogs~may therefore represent a mixed-bag population that approximately come in two flavors, i.e. ``X-ray loud'' QSO2--like and SMGs--like, which essentially indicate  different stages of quasar evolution.  
However, caution must be used before drawing any firm conclusion from these plots because of the uncertainties on the intrinsic \nh~and the AGN luminosity due to the low quality of X-ray data and the use of photometric redshifts for most of the sources considered here.\\ 

\begin{table}
\caption{The source breakdown for the \xr~sample, the {\it Red} sample, and the total sample. See Sect. 5.2 for details.}
\begin{center}
\begin{tabular}{lccc}
\hline\hline\\
\multicolumn{1}{l} {Source}&
\multicolumn{1}{c} {\xr}&
\multicolumn{1}{c} {{\it Red} Sample}&
\multicolumn{1}{c} {Total} \\
Type& Sample$^\ast$ &   & Sample \\ 
& (1) & (2) & (3) \\
 \\\hline\\ 
\nh$>10^{22}$  & 50\%  & 93\% & 66\% \\
CT AGNs        & 13\%  & 27\% & 18\% \\
QSO2s          & 43\%  & 27\% & 37\% \\
N. of Sources   & 23     & 15  & 38 \\
\hline  
\end{tabular}
\end{center}            
Percentage of objects for each source type found in the: (column 1) \xr~sample (e.g. Sect.~5); (column 2) the \xr~undetected {\it Red} sample (i.e. sources showing log(F$_{5.8}$/F$_{3.6}$) $>$ 0.2 and log(F$_{8}$/F$_{4.5}$) $>$ 0.2); (column 3) the combined (i.e. \xr+undetected {\it Red}) sample.
$^\ast$For the \xr~sample we considered the conservative values derived using the lowest value in 90\% confidence level for the \nh~parameter.
\end{table}

Fig.~6b shows the observed 2-10 keV luminosity (\lumo) as a function of  $\lambda L_{5.8 \mu m}$ for the same sources plotted in Fig.~6a.
The solid line represents the relationship obtained from the \xr~sample (i.e. Eq.~2) which can be used to estimate the intrinsic hard \xr~luminosity. The dotted and dashed line represent the observed \lumo--$\lambda L_{5.8 \mu m}$ ratio derived from Eq.~(2) in case of a column density of \nh=10$^{23}$ and \nh=10$^{24}$ cm$^{-2}$, respectively.
Downward-pointing arrows in the plot indicate the 3$\sigma$ upper limit on \lumo for the \xr~undetected sources in our sample lying in the AGN region of the IRAC color-color diagram (Fig. 1b), i.e. sources showing a power law spectrum and very red MIR colors (log(F$_{5.8}$/F$_{3.6}$) $>$ 0.2 and log(F$_{8}$/F$_{4.5}$) $>$ 0.2). Hereafter we refer to this group of 15 sources as the {\it Red} Sample.

For these 15 X-ray undetected AGNs, we converted the upper limit on the 2-10 keV flux (assuming a photon index of $\Gamma=1.9$) into the 2-10 keV observed luminosity \lumo.
The intrinsic \lum~for these sources was determined from their $\lambda L_{5.8 \mu m}$, using the relationship in Eq.~(2).
By the comparison between the observed and the intrinsic values of hard \xr~luminosity we were able to provide a basic estimate of the column density distribution in the {\it Red} Sample. We found that: (i) 93\% of these sources show a \nh~$\geq10^{22}$ cm$^{-2}$; (ii) 27\% of them show a column density \nh$>10^{24}$ cm$^{-2}$ and are Compton-Thick AGN candidates; and (iii)  QSO2s--like objects account for 27\% of the total (see Table 5).

Since 75$\%$ of the {\it Red} Sample sources have \xr~effective exposures of $\sim5$ ks, the $3\sigma$ upper limits are weakly constrained. Accordingly, the real values of the column density may likely be significantly higher than the \nh~values derived here, which should be considered as lower limits.

A handful of objects in the {\it Red} Sample, namely Nos. 2, 3 and 38, were not      
detected in X--ray medium/deep (i.e. $>$ 15 ks) observations (big downward-pointing arrows in Fig. 6b).                  
This fact makes them very interesting objects since they may likely be
more obscured than those found in the X-ray sample.                                  
In the MIR color-color diagram presented in Fig. 1b,  
these \xmogs~indeed occupy the area with the extreme red colors, i.e.  log(F$_{5.8}$/F$_{3.6}$) $>$ 0.5 and       
log(F$_{8}$/F$_{4.5}$) $>$ 0.5, that Polletta et al. (2008a) consider           
as the typical one of highly luminous and obscured quasars at $z$\simgt~1. 
In particular, we derived an \lum~$\geq$ 10$^{44}$ \ergs~for sources Nos. 2 and 3 
and \lum~$\sim$ 3 $\times$ 10$^{43}$ \ergs~for the No. 38, and the estimated \nh~value is $\geq$10$^{24}$ cm$^{-2}$ 
for all these three sources, making them very likely Compton-Thick QSO2 candidates.

Merging the results for the {\it Red} Sample with those previously found for the \xr~sample, we obtained a total sample of 38 sources for which has been possible to infer information on \nh~and hard \xr~luminosity (see Table 5).
This allows us to improve the statistics and get sounder insights on the X-ray properties\footnote{For the \xr~sample we considered the percentages derived using the lowest value in 90\% confidence level  for the \nh~parameter obtained for each source.}  of \xmogs, which basically turn out to host heavily obscured AGNs ($\sim$20\% are consistent to be Compton-thick) with high X-ray luminosity (\simgt~40\% are QSO2s). 

\section{Summary and Conclusions} 

We exploited the multiwavelength coverage of the large area SWIRE survey to
 collect  low-surface density/high luminosity (i.e. 10$^{12}$\simlt~L$_{Bol}$~ \simlt 10$^{14}$ L$_{\odot}$) objects 
on the basis of their extreme MIR/Optical flux ratio (F$_{24\mu m}$/F$_R>$2000)
 and bright MIR flux density, i.e. F$_{24\mu m}>$1.3 mJy (referred to here as \xmogs).
Cross-correlating with  publicly available \xmm/\chandra~observations of  5 SWIRE fields  we defined a final sample of 44 \xmogs~with X-ray coverage (see Table 1 and Sect. 2).
Our study has been focused on the understanding of largely unexplored X-ray spectral properties of this recently-discovered class of sources.
The main results of this paper can be summarized as follows:
\begin{itemize}

\item  All but one sources fall within the AGN region of the MIR color--color diagram proposed by Lacy et al. (2004). In particular, the vast majority of \xmogs~lies in the obscured luminous QSO locus (Polletta et al. 2008a) characterized by very red colors (Fig. 1b).\\

\item The very bulk of the redshifts, regardless of they are spectroscopic or photometric, are in the range 0.7 \simlt~$z$ \simlt~2.5 (e.g. Sect. 2.3).\\
 
\item Thanks to the large area sampled ($\sim$6 deg$^2$), our \xmogs~are on average more luminous in the MIR band by a factor of \simgt~0.5--2 orders of magnitude than their equivalents in the CDFS and COSMOS surveys (see Fig. 3b).\\

\item Twenty-three sources (the so-called X-ray sample, e.g. Sect. 4) have a positive detection either in the soft or in the hard band. The non-detection of the remaining sources in the sample can be mainly attributed to the inadequate X-ray coverage (most of the undetected sources indeed have an exposure $\leq$ 5 ks).
All the X-ray detected sources have a 2-10 keV luminosity \lum $>$ 10$^{43}$ erg s$^{-1}$, whereby they fall well within the AGN X-ray luminosity range (Fig. 4b).\\

\item Our analysis reveals that 50\% of the sources show a column density strictly larger than 10$^{22}$ \cm2. Furthermore $\sim50\%$ of them can be classified as QSO2s on the basis of their absorption properties and X-ray luminosity (i.e. \lum$>$10$^{44}$ \ergs).\\

\item At least three \xmogs~at $z >$1.4 were found to be very likely Compton-thick QSO2 candidates (e.g. Sect 5.2).
 Moreover such deeply buried AGNs may be present in a sizable fraction ($\sim30\%$, e.g. Table 5) of the X-ray undetected sources with
extreme red colors (log(F$_{5.8}$/F$_{3.6}$) $>$ 0.2 and log(F$_{8}$/F$_{4.5}$) $>$ 0.2), as suggested by their large MIR luminosity (e.g. Sect 5.2).
They therefore deserve further investigation by future properly-planned, deeper X-ray observations.\\
\end{itemize} 

\begin{figure*}
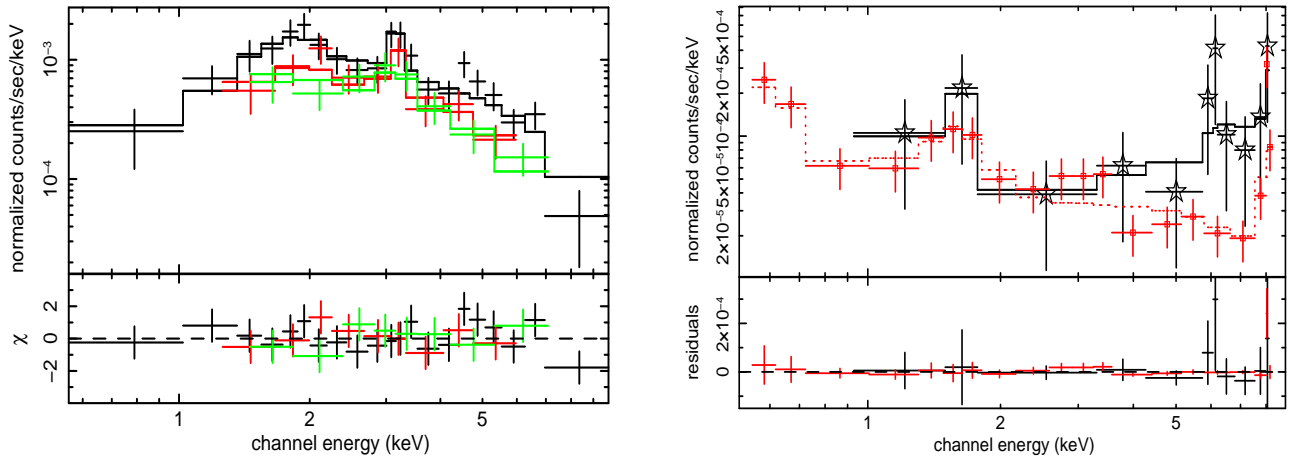

\begin{center}
\includegraphics[width=6cm,height=8cm,angle=-90]{figure_13.ps}\hspace{1cm}\includegraphics[width=6cm,height=8cm,angle=-90]{figure_14.ps}
\caption{{\it a) Left panel:} \xmm~spectrum of source No. 6 (i.e. SWIRE2 J021749.00-052306.9). The background subtracted spectra of \epic~PN (black), MOS1 (red) and MOS2 (green) are shown.{\it b) Right panel:} \xmm~spectrum of source No. 12 (i.e. SWIRE2 J022003.95-045220.4). The source spectrum (black stars) is superimposed to the background spectrum (red empty squares). The best fit models for the source (black solid line) and the background (red dotted line) are also shown.}
\end{center}
\label{fig:spettri1}
\end{figure*}  

\begin{figure*}
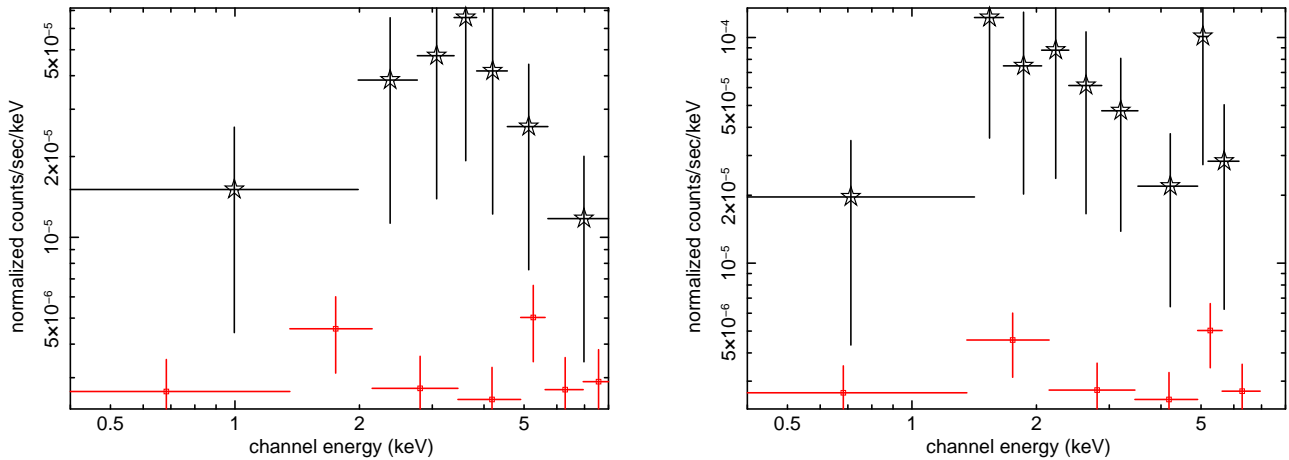
 
\label{fig:spettri2}
\begin{center}
\includegraphics[width=6cm,height=8cm,angle=-90]{figure_15.ps}\hspace{1cm}\includegraphics[width=6cm,height=8cm,angle=-90]{figure_16.ps}
\caption{{\it a) Left panel:} \chandra~spectrum of source No. 32 (i.e. SWIRE2 J104409.95+585224.7). {\it b) Right panel:} \chandra~spectrum of source No. 33 (i.e. SWIRE2 J104528.29+591326.6). Black stars (red empty squares) represent the source (background) binned counts, with 2 (15) counts per bin.}
\end{center}    
\end{figure*}  

Our work represents a further step forward in completing the census of SMBHs through cosmic epochs.
The application of our selection criteria to a large area survey has allowed to efficiently pick up  a large sample of high-$z$ sources, occupying the
bright end of the \xr~obscured AGN population.
Thus, it is a useful tool to complement the information derived by blind searches for these elusive objects performed in the pencil-beam, deep X-ray surveys
(which mostly sample the low-redshift ($z\sim$ 1), Seyfert-like AGN population) making possible the first reliable estimate of AGN "bolometric" luminosity function at high redshifts. 

 It seems probable that \xmogs~might be an early dust-embedded phase of the AGN activity, possibly associated with merger events
 (e.g. Georgantopoulos et al. 2008, Brodwin et al. 2008). Accordingly, an interesting aspect that should be examined in the next future is the FIR/sub--mm emission of these intriguing sources in order to evaluate the contribution of the star-formation activity to their large bolometric luminosity, i.e. using an approach similarly to what has been done by Pope et al. 2008a for a sample of DOGs at low F$_{24\mu m}$ levels ($\langle$F$_{24\mu m}\rangle\sim$175 $\mu$Jy).

\begin{acknowledgements}
 The authors are very grateful to the referee for the exhaustive list of helpful comments and her/his prompt response.
We thank for support the Italian Space Agency (contracts ASI-INAF I/023/05/0, ASI I/088/06/0 and PRIN MIUR 2006-02-5203).
The authors kindly thank Simonetta Puccetti, Mari Polletta and Roberto Maiolino for many useful comments. 
Based on observations obtained with XMM-Newton, an ESA science mission
with instruments and contributions directly funded by ESA Member
States and NASA. This research has made use of software provided by the Chandra X-ray Center (CXC) in the application package CIAO.
This research has made use of the NASA/IPA Extragalactic Database (NED) which is operated by the Jet Propulsion Laboratory, California Institute of Technology, under contract with the National Aeronautics and Space Administration. 

\end{acknowledgements}

\begin{appendix}
\section{Notes on individual sources}

\begin{itemize}
\item {\it Source No.~6.} This is a narrow-line quasar at $z$=0.914, as shown in on sect. 2.3.1 on the basis of a Subaru/MOIRCS spectrum.
This source is the brightest in our sample, with a 2-10 keV flux of \fhx$\sim$ 10$^{-13}$ \cgs. We fitted the \epic~data with a continuum model consisting of a power law modified by an intrinsic neutral absorption component, plus a power law with the same slope and Galactic \nh~at soft energies (\simlt~2 keV). The latter accounts for the ``soft-excess'' component, i.e. a mixture of photoionized/scattered emission superimposed upon the primary obscured continuum typically observed in most X-ray obscured AGNs (Turner et al. 1997; Piconcelli et al. 2007b; Cappi et al. 2006).
Given the redshift of this \xmog, we also included in the best-fit model a Compton reflection component with $R$ = 1 (where $R$ is the solid angle in units of 2$\pi$ subtended by the reflecting material) as commonly detected in the rest-frame 10--50 keV spectra of AGN (e.g. Risaliti 2002; Reeves et al. 2006). We obtained $\Gamma$=1.70$\pm$0.33 and \nhz $\sim$ 10.8 $\times$ \e22~\cm2, implying a 2-10 keV luminosity of \lum~ = 5.1 $\times$ 10$^{44}$ \ergs.
 The most remarkable result of our analysis is, however, the discovery of an \feka~emission line in the X-ray spectrum of this QSO2 (Fig. 7a). This line (E$_{\rm K\alpha}$=6.29$\pm$0.12 keV) is required by the data at $>$99.9\% confidence level and shows an equivalent width of EW = 590$\pm265$ eV. 
It represents one of the best examples of \feka~line in a high-$z$ QSO2s detected so far. 
   
\item {\it Sources Nos.~12, 32 and 33.} These sources (with $z$=1.443, 2.54 and 2.31 respectively) show most of their X-ray counts at $E>$ 2 keV (e.g. Figs. 7b, 8a and 8b) and they are therefore promising deeply buried  (possibly Compton-thick) QSO2 candidates. 
In particular, source No.~12 was detected with  $\sim$40(30) net counts in the 0.5-10(2-10 keV) keV band, i.e. at the limit of Cash statistic reliability for the spectral fitting procedure (Sect. 4.1). Fitting these data with a simple spectral model consisting of a power-law with fixed photon index ($\Gamma =1.9$) modified by an intrinsic absorption component, we obtained a lower limit on the column density of \nh~$> 1.4\times10^{24}$ cm$^{-2}$. The corresponding \xr~absorption-corrected luminosity is \lum$\sim 1.4\times10^{45}$ \ergs, making this \xmog~one of the best Compton-thick QSO2 candidates in our sample.

Sources Nos. 32 and 33 have only $\sim$ 20 net counts, even if they were detected in  deep ($\sim$70 ks) \chandra~exposures, and almost all of them fall in the hard X-ray band. The HR of these sources are therefore very hard ($\sim0.5$, e.g. Table 3) and suggest that they  are also highly obscured quasars. 
We derived column density values of \nh~\simgt~3 $\times$ 10$^{23}$ \cm2~for both sources (but keep in mind the large uncertainties on the HR value), confirming thus the results reported in  Weedman et al. (2006, sources A4 and A8 in their paper).
Beyond providing the spectroscopic redshifts, these authors carried out a detailed multiwavelength study for both sources that allowed to classify them as AGN-powered ULIRGs.

\item {\it Source No. 27.} This source was undetected in a $\sim$70 ks \chandra~exposure, being 
thus the only \xmog~without useful X-ray information in such a deep pointing and, possibly, the 
weakest X-ray sources in our sample. Its IRAC colors suggest that a sizable part of the 
IR emission in this source can be powered by star formation (e.g. Sect. 2.2 and Yun et al. 2008). 
According to Rowan-Robinson et al. (2008), the MIR Spectral Energy Distribution (SED) is indeed dominated by the starburst component.
This hypothesis is also supported by the large F$_{24}$/F$_{8}$ ratio ($\sim$20)  
measured for this \xmog. Such a value is similar to those reported for the 
infrared luminous galaxies at $z \sim$2 in
Yan et al. 2007, for which both deeply embedded starburst and AGN are believed 
to significantly contribute to the dust heating (e.g. Polletta et al. 2008a).
\end{itemize}
\end{appendix}


\begin{thebibliography}{}
\bibitem[Alexander et al.(2001)]{2001AJ....122.2156A} Alexander, D.~M., Brandt, W.~N., Hornschemeier, A.~E., Garmire, G.~P., Schneider, D.~P., et al. \ 2001, \aj, 122, 2156 
\bibitem[Alexander et al.(2005)]{2005Natur.434..738A} Alexander, D.~M., Smail, I., Bauer, et al.  2005, \nat, 434, 738 
\bibitem[Alexander et al.(2008)]{2008arXiv0803.0636A} Alexander, D.~M., Chary, R.~R., Pope, A., et al.\ 2008, ApJ in press, arXiv:0803.0636  
\bibitem[Alonso-Herrero et al.(2006)]{AlonsoHerrero2006} Alonso-Herrero, A., Perez-Gonz\'{a}lez, P. G., Alexander, D. M., et al., 2006, ApJ, 640, 167
\bibitem {} Antonucci, R., 1993, ARA\&A, 31, 473
\bibitem[Arnouts et al.(2007)]{2007A&A...476..137A} Arnouts, S., Walcher, C.~J., Le Fevre, O., et al.\ 2007, \aap, 476, 137 
\bibitem[Barmby et al.(2006)]{Barmby2006} Barmby, P., Alonso-Herrero, A., Donley, J. L. et al.,\ 2006, ApJ 642, 126
\bibitem {} Best, P.~N., Kauffmann, G., Heckman, T.~M, Brinchmann, J., Charlot, S., et al., \ 2005, \mnras , 362, 25 
\bibitem[Brand et al.(2006)]{2006ApJ...644..143B} Brand, K., Dey, A., Weedman, D., et al.\ 2006, \apj, 644, 143 
\bibitem[Brand et al.(2007)]{2007ApJ...663..204B} Brand, K., et al.\ 2007, \apj, 663, 204 
\bibitem[Brandt \& Hasinger(2005)]{Brandt2005} Brandt W. N., Hasinger, G., 2005, ARA\&A, 43, 827
\bibitem[Brodwin et al.(2008)]{2008ApJ...687L..65B} Brodwin, M., Dey, A., Brown, M.~J., et al.\ 2008, \apjl, 687, L65 
\bibitem {} Brown, M.~J.~I., Brand, K., Dey, A., et al. 2006, \apj, 638, 88
\bibitem[Cappi et al.(2006)]{2006A&A...446..459C} Cappi, M., et al.\ 2006, \aap, 446, 459 
\bibitem[Cardamone et al.(2008)]{2008ApJ...680..130C} Cardamone, C.~N., et al.\ 2008, \apj, 680, 130 
\bibitem[Cash(1979)]{1979ApJ...228..939C} Cash, W.\ 1979, \apj, 228, 939 
\bibitem[Cocchia et al.(2007)]{2007A&A...466...31C} Cocchia, F., et al.\ 2007, \aap, 466, 31 
\bibitem {} Comastri, A., 2004, "Multiwavelength AGN Surveys" Proceedings of the Guillermo 
Haro Conference, Edited by R. Mujica and R, Maiolino, World Scientific Publishing 
Company, Singapore, 323
\bibitem {} Cowie, L. L.; Barger, A. J.; Bautz, M. W.; Brandt, W. N.; Garmire, G. P. 2003, \apj, 584, 57
\bibitem[Croton et al.(2006)]{2006MNRAS.365...11C} Croton, D.~J., et al.\ 2006, \mnras, 365, 11 
\bibitem[Daddi et al. (2007)]{Daddi2007} Daddi, E., Alexander, D.M., Dickinson, M. et al., 2007, ApJ, 670, 173
\bibitem[Dey et al.(2008)]{2008ApJ...677..943D} Dey, A., et al.\ 2008, \apj, 677, 943 
\bibitem{}Dickey, J.~M. \& Lockman, F.~J. 1990, ARA\&A. 28, 215
\bibitem[Di Matteo et al.(2005)]{2005Natur.433..604D} Di Matteo, T., Springel, V., \& Hernquist, L.\ 2005, \nat, 433, 604 
\bibitem[Donley et al.(2007)]{Donley2007} Donley, J. L., Rieke, G. H., P\'{e}rez-Gonz\'{a}lez, P. G., Rigby, J. R., Alonso-Herrero, A., 2007, ApJ, 660, 167
\bibitem[Donley et al.(2008)]{2008ApJ...687..111D} Donley, J.~L., Rieke, G.~H., P{\'e}rez-Gonz{\'a}lez, P.~G., \& Barro, G.\ 2008, \apj, 687, 111 
\bibitem[Dwelly \& Page(2006)]{2006MNRAS.372.1755D} Dwelly, T., \& Page, M.~J.\ 2006, \mnras, 372, 1755 
\bibitem[Elvis et al. (1994)]{Elvis1994}  Elvis, M., Wilkes, B.J., McDowell, J.C. et al. 1994, ApJS, 95, 1 
\bibitem[Erlund et al.(2008)]{2008MNRAS.385L.125E} Erlund, M.~C., Fabian, A.~C., Blundell, K.~M., \& Crawford, C.~S.\ 2008, \mnras, 385, L125 
\bibitem {} Fabian, A. C. 2002, ASPC, 258, 185
\bibitem[Fabian(1999)]{1999MNRAS.308L..39F} Fabian, A.~C.\ 1999, \mnras, 308, L39 
\bibitem[Feruglio et al.(2008)]{2008A&A...488..417F} Feruglio, C., Fiore, F., La Franca, F., et al.\ 2008, \aap, 488, 417 
\bibitem[Fiore et al.(2003)]{2003A&A...409...79F} Fiore, F., Brusa, M., Cocchia, F., et al.\ 2003, \aap, 409, 79 
\bibitem[Fiore et al.(2008)]{Fiore2008} Fiore, F., Grazian, A., Santini, P., et al.\ 2008a, \apj, 672, 94 
\bibitem[Fiore et al.(2009)]{2008arXiv0810.0720F} Fiore, F., Puccetti, S., Brusa, M., et al.\ 2008b,  ApJ in press, arXiv:0810.0720 
\bibitem[Fontana et al.(2000)]{Fontana2000} Fontana, A., D'Odorico, S., Poli, et al. 2000, \aj, 120, 2206 
\bibitem[Fontana(2001)]{Fontana2001} Fontana, A.\ 2001, Astrophysics and Space Science Supplement, 277, 535 
\bibitem {} Furusawa. H., Kosugi, G., Akiyama, M., et al., 2008, ApJS, 176, 1
\bibitem[Garcet et al.(2007)]{2007A&A...474..473G} Garcet, O., Gandhi, P., Gosset, E., et al.\ 2007, \aap, 474, 473 
\bibitem[Georgantopoulos, Georgakakis \& Akylas(2007)]{Georgantopoulos2007} Georgantopoulos, I., Georgakakis, A., Akylas, A., 2007, A\&A, 466, 823
\bibitem[Georgantopoulos et al.(2008)]{2008A&A...484..671G} Georgantopoulos, I., Georgakakis, A., Rowan-Robinson, M., \& Rovilos, E.\ 2008, \aap, 484, 671 
\bibitem[Gilli et al.(2007)Gilli, Comastri \& Hasinger]{Gilli2007} Gilli, R., Comastri, A., Hasinger, G., 2007, A\&A, 463, 79
\bibitem[Granato et al.(2004)]{2004ApJ...600..580G} Granato, G.~L., De Zotti, G., Silva, L., Bressan, A., \& Danese, L.\ 2004, \apj, 600, 580 
\bibitem[Hasinger(2008)]{2008A&A...490..905H} Hasinger, G.\ 2008, \aap, 490, 905 
\bibitem[Hickox(2007)]{} Hickox, R.; Jones, C.; Forman, W. R., et al. 2007, \apj, 671, 1365,
\bibitem {} Hines, D.C., Wills, B. J., 1993, ApJ...415...82
\bibitem {} Hopkins, P.~F., Hernquist, L., Cox, T.~J., Di Matteo, T, Springel, V.,  2006, ApJS, 163, 1
\bibitem[Houck et al.(2005)]{2005ApJ...622L.105H} Houck, J.~R., Soifer, B.~T., Weedman, D., et al.\ 2005, \apjl, 622, L105 
\bibitem[Johnson et al.(2007)]{2007MNRAS.376..151J} Johnson, O., Almaini, O., Best, P.~N., \& Dunlop, J.\ 2007, \mnras, 376, 151 
\bibitem[Lacy et al.(2004)]{Lacy2004} Lacy, M., Storrie-Lombardi, L. J., Sajina, A., et al., 2004, ApJS, 154, 166
\bibitem[Lacy et al.(2007)]{2007AJ....133..186L} Lacy, M., Petric, A.~O., Sajina, A., Canalizo, G., Storrie-Lombardi, L.~J., et al. 2007, \aj, 133, 186 
\bibitem[La Franca et al.(2005)]{LaFranca2005} La Franca, F., Fiore, F., Comastri, A., et al. 2005, ApJ, 635, 864
\bibitem[Lonsdale et al.(2003)]{Lonsdale2003} Lonsdale, C.~J.,Smith, H.~E., Rowan-Robinson, M., et al.\ 2003, \pasp, 115, 897
\bibitem[Lumb et al.(2002)]{2002A&A...389...93L} Lumb, D.~H., Warwick, R.~S., Page, M., \& De Luca, A.\ 2002, \aap, 389, 93 
\bibitem[Lutz et al.(2004)]{2004A&A...418..465L} Lutz, D., Maiolino, R., Spoon, H.~W.~W., \& Moorwood, A.~F.~M.\ 2004, \aap, 418, 465 
\bibitem[Mainieri et al.(2005)]{2005A&A...437..805M} Mainieri, V., Rosati, P., Tozzi, P., et al.\ 2005, \aap, 437, 805 
\bibitem {} Maiolino, R., Comastri, A., Gilli, R., et al., 2003, MNRAS, 344, 59
\bibitem[Maiolino et al.(2006)]{2006A&A...445..457M} Maiolino, R., Mignoli, M., Pozzetti, L., et al.\ 2006, \aap, 445, 457 
\bibitem[Maiolino et al.(2007)]{2007A&A...468..979M} Maiolino, R., Shemmer, O., Imanishi, M., Netzer, H., Oliva, E., Lutz, D., \& Sturm, E.\ 2007, \aap, 468, 979 
\bibitem[Mart\'{i}nez-Sansigre et al.(2005)]{MartinezSansigre2005} Mart\'{i}nez-Sansigre, A., Rawlings, S., Lacy, M., et al., 2005, Nature, 436, 666
\bibitem[Mart{\'{\i}}nez-Sansigre et al.(2007)]{2007MNRAS.379L...6M} Mart{\'{\i}}nez-Sansigre, A., Rawlings, S., Bonfield, D.~G., et al.\ 2007, \mnras, 379, L6 
\bibitem[McMahon et al.(2001)]{2001NewAR..45...97M} McMahon, R.~G., Walton, N.~A., Irwin, M.~J., Lewis, J.~R., Bunclark, P.~S., et al.\ 2001, New Astronomy Review, 45, 97 
\bibitem[Menci et al.(2008)]{2008ApJ...686..219M} Menci, N., Fiore, F., Puccetti, S., \& Cavaliere, A.\ 2008, \apj, 686, 219 
\bibitem {} Merritt, D.,  Ferrarese, Laura, 2001, "The Central Kiloparsec of Starbursts
and AGN: The La Palma Connection",  ASP Conference Proceedings Vol. 249, 335
\bibitem[Norman et al.(2002)]{2002ApJ...571..218N} Norman, C., Hasinger, G., Giacconi, R., et al.\ 2002, \apj, 571, 218 
\bibitem[Page et al.(2004)]{2004ApJ...611L..85P} Page, M.~J., Stevens, J.~A., Ivison, R.~J., \& Carrera, F.~J.\ 2004, \apjl, 611, L85 
\bibitem[Perola et al.(2004)]{2004A&A...421..491P} Perola, G.~C., Puccetti, S., Fiore, F., et al.\ 2004, \aap, 421, 491 
\bibitem[Piconcelli et al.(2005)]{2005A&A...432...15P} Piconcelli, E., Jimenez-Bail{\'o}n, E., Guainazzi, et al. 2005 \aap, 432, 15 
\bibitem[Piconcelli et al.(2007)]{2007A&A...473...85P} Piconcelli, E., Fiore, F., Nicastro, et al. 2007a, \aap, 473, 85
\bibitem[Piconcelli et al.(2007)]{2007A&A...466..855P} Piconcelli, E., Bianchi, S., Guainazzi, M., Fiore, F., \& Chiaberge, M.\ 2007b, \aap, 466, 855 
\bibitem[Polletta et al.(2006)]{Polletta2006} Polletta, M.,  Wilkes, B. J., Siana, B., et al., 2006, ApJ, 642, 673
\bibitem[Polletta et al.(2007)]{Polletta2007} Polletta, M., Tajer, M., Maraschi, L., et al.\ 2007, \apj, 663, 81
\bibitem[Polletta et al.(2008)]{2008ApJ...675..960P} Polletta, M., Weedman, D., H{\"o}nig, S., et al.\ 2008a, \apj, 675, 960 
\bibitem[Polletta et al.(2008)]{2008A&A...492...81P} Polletta, M., Omont, A., Berta, S., et al.\ 2008b, \aap, 492, 81 
\bibitem[Pope et al.(2008)]{2008ApJ...689..127P} Pope, A., Bussmann, R. S., Dey, A., et al.\ 2008a, \apj, 689, 127 
\bibitem[Pope et al.(2008)]{2008ApJ...675.1171P} Pope, A., Chary, R. R., Alexander, D. M., et al.\ 2008b, \apj, 675, 1171 
\bibitem[Pozzi et al.(2007)]{Pozzi2007} Pozzi, F., Vignali, C., Comastri, A., et al.\ 2007, \aap, 468, 603
\bibitem[Reeves \& Turner(2000)]{2000MNRAS.316..234R} Reeves, J.~N., \& Turner, M.~J.~L.\ 2000, \mnras, 316, 234 
\bibitem {} Reeves, J. N., Fabian, A.~C., Kataoka, J., et al., 2006, AN, 327, 1079
\bibitem[Risaliti(2002)]{2002A&A...386..379R} Risaliti, G.\ 2002, \aap, 386, 379 
\bibitem[Rosati et al.(2002)]{2002ApJ...566..667R} Rosati, P., Tozzi, P., Giacconi, R., et al.\ 2002, \apj, 566, 667 
\bibitem {} Rowan-Robinson, M., Babbedge, T., Oliver, S., et al.\ 2008, MNRAS, 386, 697
\bibitem[Sajina et al.(2005)]{2005ApJ...621..256S} Sajina, A., Lacy, M., \& Scott, D.\ 2005, \apj, 621, 256 
\bibitem[Sajina et al.(2007)]{2007ApJ...664..713S} Sajina, A., Yan, L., Armus, L.\ 2007, \apj, 664, 713 
\bibitem[Seymour et al.(2008)]{2008ApJ...681L...1S} Seymour, N., Ogle, P., De Breuck, C., et al.\ 2008, \apjl, 681, L1 
\bibitem {} Sekiguchi, K., Akiyama, M., Furusawa, H., et al. 2005, Proceedings of the ESO Workshop "Multiwavelength 
mapping of galaxy formation and evolution", Edited by A. Renzini and R. Bender, 82
\bibitem[Silk \& Rees(1998)]{1998A&A...331L...1S} Silk, J., \& Rees, M.~J.\ 1998, \aap, 331, L1 
\bibitem[Simpson(2005)]{2005MNRAS.360..565S} Simpson, C.\ 2005, \mnras, 360, 565 
\bibitem[Somerville et al.(2008)]{2008MNRAS.391..481S} Somerville, R.~S., Hopkins, P.~F., Cox, T.~J., Robertson, B.~E., \& Hernquist, L.\ 2008, \mnras, 391, 481 
\bibitem[Steffen et al.(2003)]{2003ApJ...596L..23S} Steffen, A.~T., Barger, A.~J., Cowie, L.~L., Mushotzky, R.~F., \& Yang, Y.\ 2003, \apjl, 596, L23 
\bibitem[Sturm et al.(2006)]{2006ApJ...642...81S} Sturm, E., Hasinger, G., Lehmann, I., Mainieri, V., Genzel, R., et al., \ 2006, \apj, 642, 81 
\bibitem[Surace \& SWIRE(2004)]{Surace2004} Surace, J.~A., et al. 2004, Bulletin of the American Astronomical Society, 36, 1450 
\bibitem{} Tozzi, P., Gilli, R., Mainieri, V., et al., 2006, A\&A, 451, 457
\bibitem[Turner et al.(1997)]{1997ApJS..113...23T} Turner, T.~J., George, I.~M., Nandra, K., \& Mushotzky, R.~F.\ 1997, \apjs, 113, 23 
\bibitem[Ueda et al.(2003)]{2003ApJ...598..886U} Ueda, Y., Akiyama, M., Ohta, K., \& Miyaji, T.\ 2003, \apj, 598, 886 
\bibitem{} Vignali et al. 2008 submitted
\bibitem[Weedman et al.(2006)]{2006ApJ...638..613W} Weedman, D.~W., Le Floc'h, E., Higdon, S.~J.~U., et al.\ 2006, \apj, 638, 613 
\bibitem[Worsley et al.(2005)]{2005MNRAS.357.1281W} Worsley, M.~A., Fabian, A.~C., Bauer, F.~E., et al.\ 2005, \mnras, 357, 1281 
\bibitem {} Yan, L., Sajina, A., Fadda, D., et al., 2007, ApJ, 658, 778
\bibitem {} Yan, L., Chary, R., Armus, L., et al., 2005, ApJ, 628, 604
\bibitem[Yun et al.(2008)]{2008MNRAS.389..333Y} Yun, M.~S., Aretxaga, I., Ashby, M.~L.~N., et al.\ 2008, \mnras, 389, 333 
\end{thebibliography}
\end{document}